\documentclass[twocolumn,showpacs,preprintnumbers,amsmath,amssymb,pra,superscriptaddress]{revtex4}
\usepackage[latin1]{inputenc}
\usepackage{dcolumn,color}
\usepackage{bm}
\usepackage{graphicx}
\usepackage[english]{babel}
\usepackage[bbgreekl]{mathbbol}
\usepackage{bbm}
\usepackage{stmaryrd}

\DeclareMathOperator{\Rea}{Re}

\def\bbm[#1]{\mbox{\boldmath $#1$}}
\newcommand{\ket}[1]{\displaystyle{|#1\rangle}}
\newcommand{\bra}[1]{\displaystyle{\langle #1|}}
\newcommand{\TE}{\text{TE}}
\newcommand{\TM}{\text{TM}}

\makeatletter
\def\amsbb{\use@mathgroup\M@U\symAMSb}
\makeatother

\usepackage{hyperref}
\hypersetup{colorlinks,
linkcolor=black,
filecolor=green,
urlcolor=blue,
citecolor=black}

\begin{document}

\title{Casimir interaction between a sphere and a grating}

\author{Riccardo Messina}
\affiliation{Laboratoire Charles Coulomb, UMR 5221 Universit\'{e} de Montpellier and CNRS, F- 34095 Montpellier, France}
\author{Paulo A. Maia Neto}
\affiliation{Instituto de F\'{i}sica, Universidade Federal do Rio de Janeiro, Rio de Janeiro, RJ 21941-972, Brazil}
\author{Brahim Guizal}
\affiliation{Laboratoire Charles Coulomb, UMR 5221 Universit\'{e} de Montpellier and CNRS, F- 34095 Montpellier, France}
\author{Mauro Antezza}
\affiliation{Laboratoire Charles Coulomb, UMR 5221 Universit\'{e} de Montpellier and CNRS, F- 34095 Montpellier, France}
\affiliation{Institut Universitaire de France, 1 rue Descartes, F-75231 Paris Cedex 05, France}

\date{\today}

\begin{abstract}
We derive the explicit expression for the Casimir energy between a sphere and a 1D grating, in terms of the sphere and grating reflection matrices, and valid for arbitrary materials, sphere radius, and grating geometric parameters. We then numerically calculate the Casimir energy between a metallic (gold) sphere and a dielectric (fused silica) lamellar grating at room temperature, and explore its dependence on the sphere radius, grating-sphere separation, and lateral displacement. We quantitatively investigate the geometrical dependence of the interaction, which is sensitive to the grating height and filling factor, and show how the sphere can be used as a local sensor of the Casimir force geometric features. To this purpose we mostly concentrate on separations and sphere radii of the same order of the grating parameters (here of the order of one micrometer). We also investigate the lateral component of the Casimir force, resulting from the absence of translational invariance. We compare our results with those obtained within the proximity force approximation (PFA). When applied to the sphere only, PFA overestimates the strength of the attractive interaction, and we find that the discrepancy is larger in the sphere-grating than in the sphere-plane geometry. On the other hand, when PFA is applied to both sphere and grating, it provides a better estimate of the exact results, simply because the effect of a single grating is underestimated, thus leading to a partial compensation of errors.
\end{abstract}


\maketitle

\section{Introduction}

Over the last decades, a remarkable progress in the control of the Casimir interaction \cite{Casimir} between material surfaces has been achieved \cite{book}, paving the way for technological applications in the nanoscale \cite{nanoscale1,nanoscale2}. In that respect, nanostructured surfaces are particularly promising, for they allow to tailor the Casimir interaction, leading for instance to a strong force reduction in the case of a metallic nanostructured grating with a period below the plasma wavelength \cite{IntravaiaNatComm14}. In addition, they give rise to a lateral Casimir force \cite{Chen2002, Chiu2010,BenderPRX14} and a Casimir torque \cite{Rodrigues2006,Cavero-Pelaez2008} that could be useful for the design of novel actuation schemes in nanoelectromechanical systems (NEMS).

From a more fundamental point of view, nanostructured surfaces are ideally suited to highlight the non-trivial geometry dependence of the Casimir interaction. Since dispersive interactions are non-additive \cite{Milonni94}, it is not possible to build up the Casimir interaction between material surfaces from the elementary van der Waals or Casimir-Polder interaction between their atomic constituents. For some geometries, such pairwise summation approach is unable to predict even the attractive or repulsive nature of the interaction \cite{Levin2010}.

The proximity force approximation (PFA) \cite{DeriaginQuartRev68} provides an alternative approach valid in the limit of short distances and large curvature radii. Within PFA, the Casimir interaction energy is approximated by the corresponding result for parallel planes after averaging over the local distances. Recent advances provide insights on the validity of PFA in some particular conditions \cite{Fosco11,Teo11,Bimonte12,Bimonte12b,Canaguier12,Fosco15}. Nonetheless PFA still remains in general a non-controlled approximation, especially with non-gentle geometries, and to have some quantitative and qualitative meaning it should be compared with the exact calculation.

More recently, the development of the scattering approach \cite{LambrechtNewJPhys06,EmigPRL07} led to the derivation of exact results for the sphere-plane configurations, as well as for a variety of non-trivial geometries, allowing to assess the accuracy of different approximation methods. When applied to nanostructured gratings, the key ingredient is the evaluation of the grating reflection matrix. The simplest derivation is based on a perturbative approximation, valid for small grating amplitudes \cite{Rodrigues2006B1,Rodrigues2006B2,MessinaPRA09}. The full exact calculation, involving a numerical computation of the reflection matrix elements, was applied to configurations at \cite{Marachevsky2008,Davids2010,Contreras-ReyesPRA10,Intravaia2012,Guerout2013} or out \cite{NotoPRA14} of thermal equilibrium.

Comparison with the exact results shows that PFA underestimates the attractive force in the case of a single grating interacting with a planar surface \cite{Davids2010,Intravaia2012}, while it overestimates the interaction between two gratings \cite{Rodrigues2006B1,Rodrigues2006B2,Marachevsky2008}. All these models considered the geometry of a planar grating interacting either with a planar surface or a second lamellar grating. However, the geometry of experimental relevance involves a spherical surface instead, either smooth \cite{IntravaiaNatComm14,Chan2008,Chan2010} or with an imprinted grating \cite{Chen2002,Chiu2010,Banishev2013}. In the presence of gratings, up to now the sphere has been always analyzed with the help of PFA only. Hence an exact calculation taking into account both the grating and the sphere geometry is missing. Even though the employed distance-to-radius aspect ratios, typically in the range $10^{-2}\div10^{-3},$ suggest that PFA should provide an accurate description of the experimental conditions, it is still important to evaluate the validity of PFA by comparing its results with an exact calculation. This is particularly important in connection with the experiment reported in \cite{IntravaiaNatComm14}, where a significant disagreement between PFA theory and experimental data was found.

In this paper, we compute the exact interaction energy between a dielectric grating and a metallic sphere, without approximations, by considering the full scattering matrices of both sphere and grating in the framework of the scattering approach.
Although we are not able to numerically compute for the very small aspect ratios probed experimentally, our results indicate that PFA provides indeed a less accurate description of the sphere curvature in the sphere-grating geometry than in the sphere-plane one. Moreover, applying PFA to the sphere overestimates the attractive interaction, although by a margin which is too small to explain the theory-experiment discrepancy found in \cite{IntravaiaNatComm14}.

By taking the spherical geometry fully into account, we are also able to discuss the Casimir energy dependence on the sphere lateral position. Small spheres act as local probes of the field fluctuations, which are not translational-invariant as a consequence of the grating profile \cite{Dalvit2008,Dobrich2008,MessinaPRA09,Contreras-ReyesPRA10}. As in the parallel-grating setup~\cite{Chen2002, Chiu2010}, a lateral Casimir force appears for small and medium sized spheres.

This paper is organized as follows. In Sec.~\ref{SecDef}, we present our notations and definitions and discuss our theoretical method. Sec.~\ref{SecScatt} is devoted to the evaluation of the grating and sphere scattering matrices, which are the key ingredients of the scattering method. In Sec.~\ref{SecNum}, we present our numerical results and physical discussions. Concluding remarks are presented in Sec.~\ref{SecConcl}. The Appendix contains a detailed derivation of all ingredients required in the analytical derivation.

\section{Physical system and definitions}\label{SecDef}

\begin{figure}[h]\centering
\includegraphics[height=7.5cm]{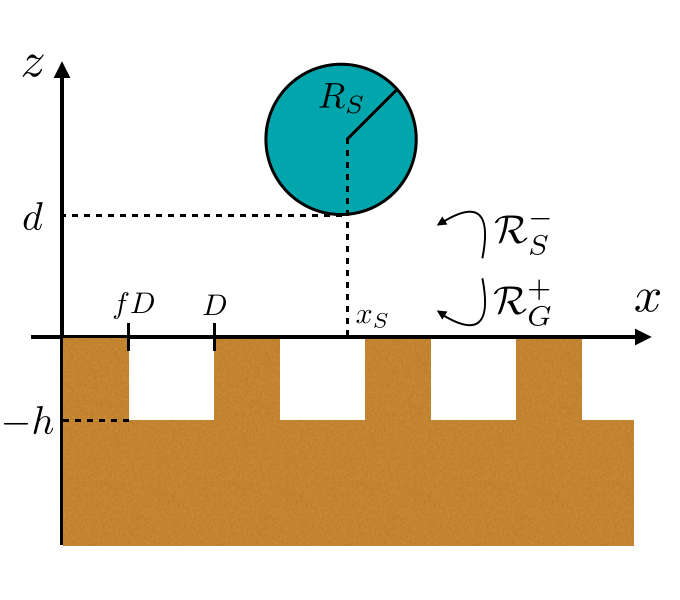}
\caption{Geometry of the system. A sphere of radius $R_S$ is placed at distance $d$ from the upper plane of a 1D grating, periodic along the $x$ axis with period $D$. Moreover, $f$ is the filling factor of the grating, $h$ its depth, while $x_S$ is the $x$ coordinate of the sphere center. $\mathcal{R}_S^-$ and $\mathcal{R}_G^+$ are the reflection matrices involved in the calculation of the Casimir interaction (see text).}\label{Geometry}\end{figure}

We consider the Casimir interaction between a sphere and a 1D lamellar grating, as depicted in Fig.~\ref{Geometry}. The sphere has radius $R_S$, it is placed at distance $d$ from the highest part of a grating of period $D$ and depth $h$. The unit cell of the grating is divided in two zones: one, going from $x=0$ to $x=fD$ ($f$ being the filling factor) is filled with a given arbitrary material; the remaining region from $x=fD$ to $x=D$ is empty. This periodic structure is placed on top of a semi-infinite substrate ($z<-h$) made by the same material of the grating (the case of substrate of finite thickness is a trivial extension resulting in a minor modification of the grating reflection matrix, and can be found in \cite{NotoPRA14}).

While the following analytical derivation is valid for arbitrary sphere and grating dielectric properties, numerical investigations will involve a metallic gold (Au) sphere and a dielectric fused silica (SiO$_2$) grating. We address the calculation of the Casimir free energy assuming that the system is at thermal equilibrium at temperature $T$. In particular, throughout all this work, the temperature is assumed to be $T=300\,$K, implying a natural energy scale given by the
thermal energy $k_BT=25.9\,$meV. Among several theoretical approaches, the Casimir free energy can be calculated by means of the scattering-matrix method. This technique, based on the description of each interacting body by its individual reflection and transmission operators \cite{JaekelReynaud91}, has been largely developed and exploited to discuss Casimir forces both at \cite{LambrechtNewJPhys06,EmigPRL07,RahiPRD09,Rodrigues2006B1,Rodrigues2006B2,MaiaNetoPRA08,CanaguierDurandPRL09,CanaguierDurandPRL10,CanaguierDurandPRA10,EmigJStatMech08} and out \cite{BimontePRA09,MessinaEurophysLett11,MessinaPRA11,MessinaPRA14} of thermal equilibrium. The equilibrium Casimir free energy is written as a sum over the Matsubara frequencies
\begin{equation}\xi_n=\frac{2\pi nk_BT}{\hbar},\qquad n=0,1,2,...\end{equation}
as follows:
\begin{equation}\label{ET}{\cal F}=k_BT{\sum_n}'\log\det(\mathbb{1}-\mathcal{M}(\text{i}\xi_n)),\end{equation}
where the prime stands for an additional factor $1/2$ when accounting for the zero-frequency ($n=0$) contribution. The round-trip operator $\mathcal{M}$ is defined as
\begin{equation}\label{M}\mathcal{M}=\mathcal{R}_S^-e^{-\mathcal{K}(d+R_S)}\mathcal{R}_G^+e^{-\mathcal{K}(d+R_S)}\end{equation}
where $\mathcal{R}_G^+$ ($\mathcal{R}_S^-$) is the operator accounting for the reflection of waves propagating along the negative (positive) $z-$direction by the grating (sphere) (see Fig.~\ref{Geometry}), and $e^{-\mathcal{K}(d+R_S)}$ is the translation operator, where $\mathcal{K}$ is diagonal in the plane-wave basis with matrix elements equal to the imaginary part of the wavevector $z-$component (see Eq.~\eqref{kk} below for the explicit expression).

The calculation of the sphere-grating Casimir interaction from Eq.~\eqref{ET} is challenging because this configuration mixes the planar and spherical symmetries. In fact, the descriptions of the grating and sphere reflection operators are easily achieved, when taken separately, in the plane-wave and spherical-wave bases, respectively. As in the plane-sphere configuration, our method will then rely on the simultaneous use of these two bases, so as to take full advantage of both symmetries \cite{MaiaNetoPRA08,CanaguierDurandPRA10,CanaguierDurandPRL09,CanaguierDurandPRL10}. This approach requires, as a preliminary step, the derivation of the change of basis matrix elements that allow us to combine the two bases when evaluating the Casimir energy \eqref{ET}.

Both bases are defined for a fixed frequency $\omega,$ which is later replaced by $i\xi_n.$ Each mode $(\mathbf{k},p,\phi)$ in the plane-wave basis is then defined by the wavevector component parallel to the $xy$ plane $\mathbf{k},$ the polarization $p=1,2$ corresponding to transverse electric (TE) or transverse magnetic (TM), respectively, and the parameter $\phi=\pm 1$ fixing the propagation direction along the $z$ axis. Thus, the wavevector $z$ component is $\phi k_z,$ where $k_z$ is a dependent variable defined as
\begin{equation}k_z=\sqrt{\frac{\omega^2}{c^2}-\mathbf{k}^2}.\end{equation}
The complete wavevector $\mathbf{K}$ reads
\begin{equation}\mathbf{K}^\phi=(\mathbf{k},\phi k_z)=(k_x,k_y,\phi k_z),\end{equation}
and has modulus $K=\omega/c$.

The spherical-wave modes $(\ell,m,P,s)$ are transverse vector fields that are eigenfunctions of the total angular momentum $\mathbf{L}^2$ and its $z$-component $L_z$ with eigenvalues $\ell(\ell+1)\hbar^2$ and $m\hbar$, respectively ($\ell=1,2,...,m=-\ell,...,\ell$). $P$ denotes the spherical polarization, taking the values $P=\text{E}$ (electric multipole) and $P=\text{M}$ (magnetic multipole). The fourth index $s$ takes values $s=$\,reg and $s=$\,out and denotes respectively regular modes (where the radial dependence of the mode is given by the Bessel function $j_\ell$, see Appendix for more details) and outgoing modes
(given in terms of the spherical Hankel function of the first kind $h^{(1)}_\ell$).

In the Appendix, we derive the change of basis matrix elements we will need in the following. We find
\begin{equation}\label{Proj1}\begin{split}\bra{\ell,m,\text{E},\text{reg}}&\mathbf{k},\text{TE},\phi\rangle=-i\bra{\ell,m,\text{M},\text{reg}}\mathbf{k},\text{TM},\phi\rangle\\
&=-\frac{4\pi me^{-im\varphi_\mathbf{k}}}{\sqrt{\ell(\ell+1)}\sin\theta^\phi_\mathbf{k}}Y_{\ell m}(\theta^\phi_\mathbf{k},0),\\
\bra{\ell,m,\text{E},\text{reg}}&\mathbf{k},\text{TM},\phi\rangle=i\bra{\ell,m,\text{M},\text{reg}}\mathbf{k},\text{TE},\phi\rangle\\
&=-\frac{4\pi ie^{-im\varphi_\mathbf{k}}}{\sqrt{\ell(\ell+1)}}\frac{\partial Y_{\ell m}(\theta^\phi_\mathbf{k},0)}{\partial\theta},\end{split}\end{equation}
and
\begin{equation}\label{Proj2}\begin{split}
\bra{\mathbf{k},\text{TE},\phi}&\ell,m,\text{E},\text{out}\rangle=i\bra{\mathbf{k},\text{TM},\phi}\ell,m,\text{M},\text{out}\rangle\\
&=-\frac{2\pi m}{Kk_z\sqrt{\ell(\ell+1)}\sin\theta^\phi_\mathbf{k}}Y_{\ell m}(\theta^\phi_\mathbf{k},\varphi_\mathbf{k}),\\
\bra{\mathbf{k},\text{TM},\phi}&\ell,m,\text{E},\text{out}\rangle=-i\bra{\mathbf{k},\text{TE},\phi}\ell,m,\text{M},\text{out}\rangle\\
&=\frac{2\pi i}{Kk_z\sqrt{\ell(\ell+1)}}\frac{\partial Y_{\ell m}(\theta^\phi_\mathbf{k},\varphi_\mathbf{k})}{\partial\theta},
\end{split}\end{equation}
where $\theta^\phi_{\bf k}$ and $\varphi_\mathbf{k}$ are the spherical angles defining the direction of $\mathbf{K}^\phi$ and $Y_{\ell m}$ are the spherical harmonics.

It is worth stressing that Eqs.~\eqref{Proj1} and \eqref{Proj2} differ from the ones used in \cite{CanaguierDurandPRA10} and related works. Apart from a slightly different definition of the spherical modes, Eqs.~\eqref{Proj1} and \eqref{Proj2} result in a sphere scattering matrix which satisfies the reciprocity relations and gives the correct limit for small radius (see Secs.~\ref{RSPW}, \ref{recrel} and \ref{atom} in Appendix for more details).

\section{Scattering matrices}\label{SecScatt}

In this section, we consider the grating and sphere scattering matrices, and then derive an explicit expression for the matrix representing the round-trip operator $\cal M$, given by \eqref{M}, in the spherical-wave basis. This result allows us to numerically compute the Casimir free energy \eqref{ET} in Sec.~\ref{SecNum}

\subsection{1D-grating scattering matrix}

In order to take into account the periodicity along the $x$ axis, we employ a mode decomposition where the wavevector component $k_x$ is replaced by the new mode parameter
\begin{equation}k_{x,n}=k_x+\frac{2\pi}{D}n,\end{equation}
with $k_x$ taking values in the first Brillouin zone $[-\pi/D,\pi/D]$ and $n$ assuming all integer values. A given plane-wave mode is then written as $|k_x,k_y,n,p,\phi\rangle$. The reflection upon the grating conserves both $k_y$ (because of translational invariance) and $k_x$ in the first Brillouin zone (as a consequence of periodicity). Thus, the generic matrix element of the grating reflection operator satisfies the following relation (for a given frequency $\omega$)
\begin{equation}\label{DefRG}\begin{split}\bra{k_x,k_y,n,p,+}&\mathcal{R}_G^+\ket{k'_x,k'_y,n',p',-}\\
&=(2\pi)^2\delta(k_x-k'_x)\delta(k_y-k'_y)\\
&\,\times\bra{n,p}\mathcal{R}_G^+(k_x,k_y)\ket{n',p'}.\end{split}\end{equation}
We are actually left with a matrix $\mathcal{R}_G^+(k_x,k_y)$ operating on the discrete indexes $n\in\mathbb{Z}$ and $p=1,2$. The index $n$ can still take an infinite number of values, and the number of possible diffraction orders has to be truncated.

In the numerical application we will consider a 1D lamellar dielectric grating, and use the Fourier Modal Method (FMM) to analyze the electromagnetic reflection and hence derive its scattering matrix. The explicit form of such a matrix, together with a detailed discussion on the numerical convergence can be found in \cite{NotoPRA14}.

\subsection{Sphere scattering matrix}

The scattering upon a homogeneous sphere (permittivity $\epsilon$) conserves the angular momentum variables $\ell$ and $m$ as well as the polarization $P$. It is thus convenient to define
\begin{equation}\label{DefRS}\bra{\ell,m,P,s}\mathcal{R}_S^-\ket{\ell',m',P',s'}=\delta_{\ell \ell'}\delta_{mm'}\delta_{PP'}\delta_{s,\text{out}}\delta_{s',\text{reg}}r_{\ell P}.\end{equation}
This expression
represents the scattering of an incident spherical mode, regular at the origin, into an outgoing spherical wave. The scattering amplitudes $r_{\ell P}(\omega)$ are independent of $m$ by symmetry. They correspond, apart from the sign, to the well-known Mie coefficients \cite{Bohren} $a_{\ell}= -r_{\ell E}$ and $b_{\ell}= -r_{\ell M}$ for a homogenous sphere. Here we provide the explicit expressions on the imaginary frequency axis, in terms of the modified Bessel functions \cite{Abramowitz} evaluated at the ``size parameter'' variables $x=\xi R_S/c$ and $n x,$ with $n(i\xi)=\sqrt{\varepsilon(i\xi)}$ representing the sphere refractive index:
\begin{equation}\label{DefrlSi}\begin{split}r_{\ell\text{E}}(i\xi)&=(-1)^{\ell}\frac{\pi}{2}\frac{\varepsilon(i\xi)S_\ell^{(a)}(x)-S_\ell^{(b)}(x)}
{\varepsilon(i\xi)S_\ell^{(c)}(x)-S_\ell^{(d)}(x)},\\ r_{\ell\text{M}}(i\xi)&=(-1)^{\ell}\frac{\pi}{2}\frac{S_\ell^{(a)}(x)-S_\ell^{(b)}(x)}{S_\ell^{(c)}(x)-S_\ell^{(d)}(x)},\end{split}\end{equation}
\begin{equation}\begin{split}S_\ell^{(a)}(x)&=I_{\ell+\frac{1}{2}}(n(ix)x)\bigl[xI_{\ell-\frac{1}{2}}(x)-\ell I_{\ell+\frac{1}{2}}(x)\bigr],\\
S_\ell^{(b)}(x)&=I_{\ell+\frac{1}{2}}(x)\bigl[n(ix)xI_{\ell-\frac{1}{2}}(n(ix)x)-\ell I_{\ell+\frac{1}{2}}(n(ix)x)\bigr],\\
S_\ell^{(c)}(x)&=I_{\ell+\frac{1}{2}}(n(ix)x)\bigl[-xK_{\ell-\frac{1}{2}}(x)-\ell K_{\ell+\frac{1}{2}}(x)\bigr],\\
S_\ell^{(d)}(x)&=K_{\ell+\frac{1}{2}}(x)\bigl[n(ix)xI_{\ell-\frac{1}{2}}(n(ix)x)-\ell I_{\ell+\frac{1}{2}}(n(ix)x)\bigr].\end{split}\end{equation}

\vspace{.2cm}
\subsection{Round-trip scattering matrix}

In order to obtain an explicit expression for the matrix elements of the round-trip operator $\mathcal{M}$ given by \eqref{M}, we still need the matrix elements of the $\mathcal{K}$ operator, which read (for a given imaginary frequency $\omega=i\xi$)
 \begin{equation}\label{kk}\begin{split}&\bra{k_x,k_y,n,p,\phi}\mathcal{K}\ket{k'_x,k'_y,n',p',\phi}\\
&=(2\pi)^2\delta(k_x-k'_x)\delta(k_y-k'_y)\delta_{nn' }\delta_{pp' }\kappa_n,\end{split}\end{equation}
where
\begin{equation}\kappa_n=\sqrt{\frac{\xi^2}{c^2}+k_{x,n}^2+k_y^2}.\end{equation}

To compute the determinant in \eqref{ET} using the spherical-wave basis, we first remark from Eq.~\eqref{DefRS} that the operator $\mathcal{M},$
representing a closed round-trip \textit{loop}, only has matrix elements between outgoing modes $s=$\,out.

We insert twice the spectral decomposition of the identity operator (closure relation) in terms of the plane-wave basis
\begin{equation}\mathbb{I}=\sum_{n, p,\phi}\int_{-\frac{\pi}{D}}^{\frac{\pi}{D}}\frac{dk_x}{2\pi}\int_{-\infty}^{+\infty}\frac{dk_y}{2\pi}|\mathbf{k},n,p,\phi\rangle \langle\mathbf{k},n,p,\phi|,\end{equation}
into the r.-h.-s. of \eqref{M}
and use {Eqs.~\eqref{DefRG}, \eqref{DefRS} and \eqref{kk} } to find
\begin{equation}\label{MatEl}\begin{split}\bra{\ell_1,m_1,P_1,\text{out}}&\mathcal{M}\ket{\ell_2,m_2,P_2,\text{out}}=\\
&=\int_{-\frac{\pi}{D}}^{\frac{\pi}{D}}\frac{dk_x}{2\pi}\int_{-\infty}^{+\infty}\frac{dk_y}{2\pi}\sum_{n,n'}\sum_{p,p'}r_{\ell_1P_1}\\
&\,\times\bra{\ell_1,m_1,P_1,\text{reg}}\mathbf{k},n',p',+\rangle\\
&\,\times e^{-\kappa_{n'}(d+R_S)}\bra{n',p'}\mathcal{R}_G^+(k_x,k_y)\ket{n,p}\\
&\,\times e^{-\kappa_{n}(d+R_S)}\langle\mathbf{k},n,p,-\ket{\ell_2,m_2,P_2,\text{out}}.\end{split}\end{equation}
The matrix element \eqref{MatEl} represents the scattering amplitude for going from a spherical mode $\ket{\ell_2,m_2,P_2,\text{out}}$ to a different mode $\ket{\ell_1,m_1,P_1,\text{out}}$ after one complete round-trip propagation between the two interacting bodies. It can be easily interpreted if read from right to left. The mode $\ket{\ell_2,m_2,P_2,\text{out}}$, coming from the sphere, is first decomposed into plane waves that travel from the sphere to the grating. After multiplication by the translation factor {(free space propagation),} each plane wave is reflected by the grating, changing in general its diffraction order and polarization, and then propagates back towards the sphere with a new wavevector value and multiplied by the appropriate translation factor. This plane wave is now decomposed into spherical regular modes, being plane waves regular at the origin. Since the scattering by the sphere conserves angular momentum, the final transition amplitude to the outgoing mode $\ket{\ell_1,m_1,P_1,\text{out}}$ corresponds to the multiplication by the Mie coefficient $r_{\ell_1P_1}$. The total amplitude is obtained by adding over all plane-wave intermediate modes.

Each matrix element of $\mathcal{M}$ is then a double integral, with respect to $k_x$ and $k_y$, of a matrix product in the vector space
spanned by the integers $(n,p)$. All expressions appearing in \eqref{MatEl} were explicitly given in Secs.~\ref{SecDef}.C and \ref{SecScatt}.B, except for the grating reflection matrix, whose evaluation is discussed in detail in Ref.~\cite{NotoPRA14}. We now have all the ingredients needed to compute the Casimir free energy.

\section{Numerical results}\label{SecNum}

We will present in this section numerical results for the Casimir energy between a sphere and a grating for different configurations. First of all, we want to stress that the sphere-grating calculation proves to be much more challenging than the one for a sphere in front of a plane. The first obvious reason is that, whereas the reflection matrix of a plane is diagonal with respect to wavevectors and known analytically (for instance it is simply given in terms of the Fresnel coefficients in the case of a homogeneous medium), the situation is different in the case of a grating, where a numerical method has to be employed to numerically evaluate a non-diagonal matrix linking wavevectors differing in their $x$ component by multiples of $\frac{2\pi}{D}$.

Moreover, the grating breaks the rotational symmetry around the $z$-axis, so that the round-trip matrix $\cal M$ is no longer block-diagonal in $m$, representing the angular-momentum $z$ component. Whereas the contributions from different values of $m$ can be calculated independently in the plane-sphere geometry \cite{MaiaNetoPRA08}, in the sphere-grating case the full non-diagonal matrix coupling different values of $m$ has to be considered.

When computing the determinant of the round-trip matrix $\cal M,$ we must truncate the number of multipoles up to a given maximum value $\ell_{\rm max}.$
As in the plane-sphere case, the required $\ell_{\rm max}$ is proportional to the radius to distance ratio $R_S/d$. Coherently with previous results \cite{MaiaNetoPRA08}, we have taken $\ell_\text{max}=4d/R_S$, in order to achieve a precision of the order of 1\%. We also need to truncate the sums over the diffraction orders $n$ and $n'$ in \eqref{MatEl} when computing the grating reflection using the FMM method. We used the same reasoning presented in \cite{NotoPRA14}, and the maximum number of diffraction orders that need to be taken into account mainly depends on the distance $d$ and on the grating period $D$.

Before moving to the results, let us state our choice for the materials of the sphere and the grating. The sphere is made of gold, described using a Drude model
\begin{equation}\varepsilon_S(i\xi)=1+\frac{\omega_P^2}{\xi(\xi+\gamma)},\end{equation}
where the plasma frequency and the dissipation rate are respectively equal to $\omega_P=9\,$eV and $\gamma=35\,$meV. Concerning the grating, it is made of fused silica, for which we used optical data \cite{Palik98} and a suitable Kramers-Kronig relation in order to obtain the permittivity function along the imaginary axis \cite{Lambrecht2000}.

\subsection{Casimir energy for different filling factors}

We start our numerical discussion presenting the dependence of the Casimir energy on the sphere-grating distance $d$. The sphere radius is $R_S=5\,\mu$m and for each configuration the sphere center is aligned with the center of the higher part of the grating. The grating has depth $h=500\,$nm and we consider four different filling factors $f=0.3,0.5,0.7,1.0$. The last case corresponds of course to a plane, and we have verified the agreement with an independent sphere-plane calculation.

\begin{figure}[h]\centering
\includegraphics[height=7.5cm]{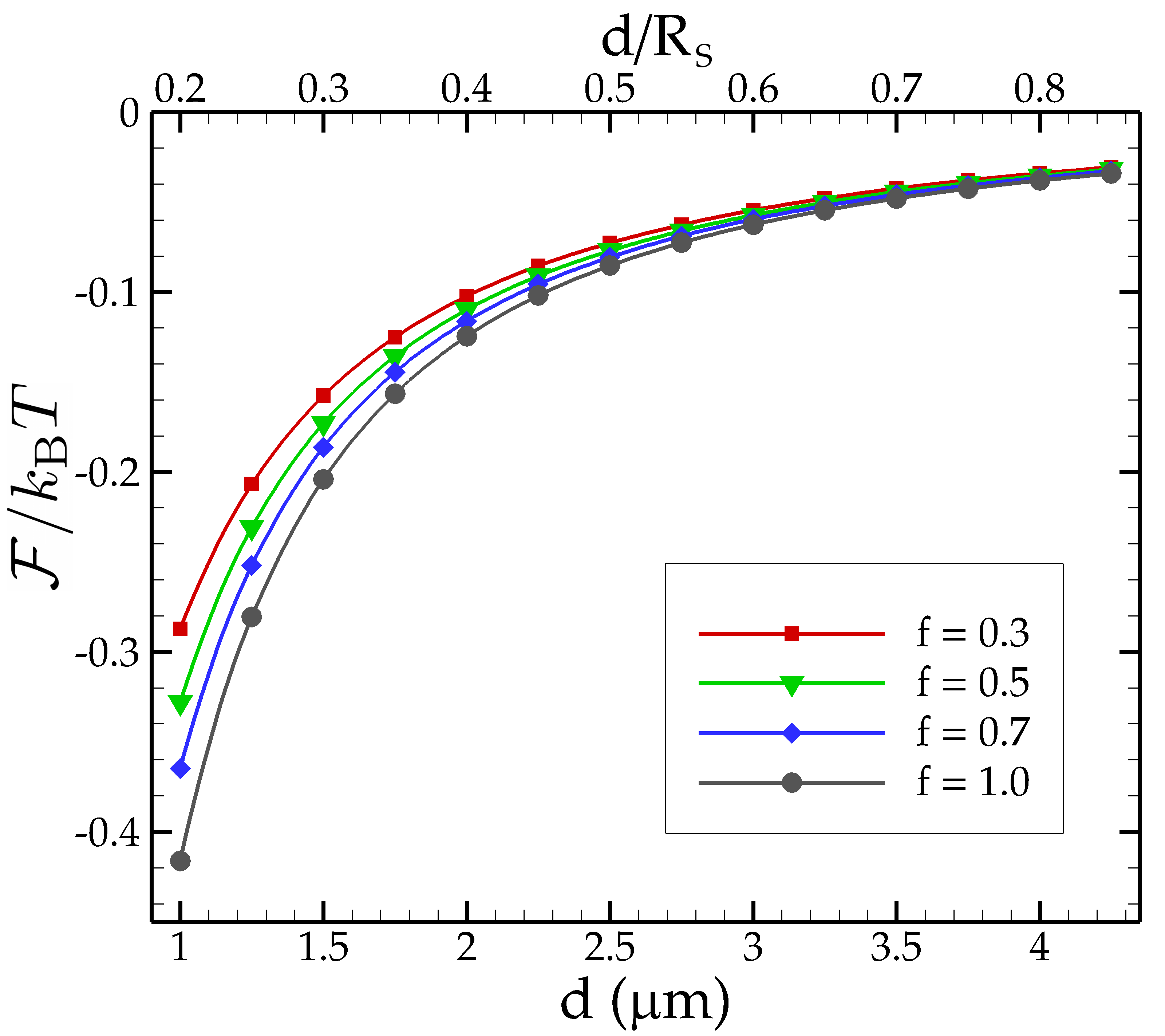}
\caption{Casimir interaction energy (in units of $k_BT=25.9\,$meV) between a gold sphere (radius $R_S=5\,\mu$m) and a grating made of fused silica, as a function of the distance $d$. The grating has period $D=1\,\mu$m, depth $h=500\,$nm and four possible filling factors: $f=0.3,0.5,0.7,1$, the last one corresponding to a sphere-plane configuration. The points are obtained numerically and the solid lines are interpolations.}\label{FigEf}\end{figure}

In Fig.~\ref{FigEf} we show the Casimir energy $\cal F$ as a function of $d$. We took distances ranging from $d=1\,\mu$m to $4.25\,\mu$m, corresponding to ratios $d/R_S$ in the range [0.2,0.85]. We observe that the energy has a monotonic behavior with respect to $d$ for any filling factor. Moreover, not surprisingly, for a given distance $d$, the energy is also an increasing function of $f$. The same monotonic behavior with respect to $f$ was also observed in the case of two gratings \cite{NotoPRA14}.

\subsubsection{Comparison with the proximity force approximation}

It is now instructive to compare the exact results with the ones obtained in the context of the Proximity Force Approximation (PFA) \cite{DeriaginQuartRev68}. This approach is widely used to deal with complex geometries and in particular in the case of plane-sphere interactions. In this simple configuration it corresponds to the decomposition of the sphere into concentric hollow cylinders extending in $z$ from the sphere to $+\infty$, orthogonal to the plane, and of infinitesimal thickness. Each one contributes to the Casimir energy with the product of its frontal area times the Casimir energy per unit of surface between two half spaces at the corresponding distance. The final result is the integral of these contributions. Of course this approach does not take into account the well known non-additivity of Casimir interactions \cite{Milonni94} and for this reason its precision cannot be assessed, except by comparison with exact results taking the sphere curvature fully into account.

\begin{figure}[h]\centering
\includegraphics[height=7.5cm]{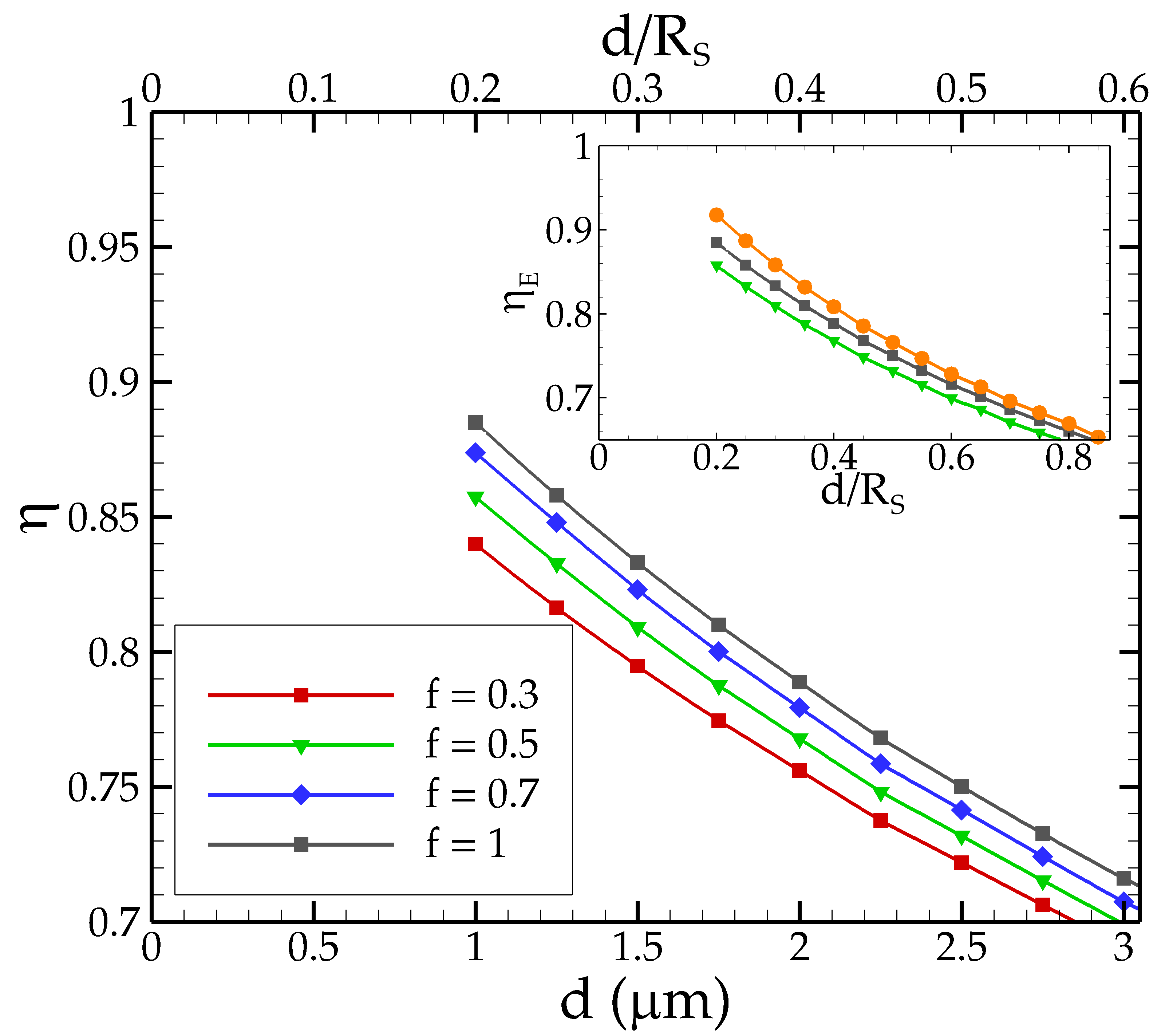}
\caption{Main part of the plot: ratio between the exact Casimir energy and the PFA result \eqref{PFA1} as function of distance for four different values of the filling factor $f$ (see legend). In the inset, the green triangles and grey squares are the same as in the main part of the figure, while the orange circles {\color{black}represent} the ratio between the exact result and the double PFA \eqref{PFA2} for $f=0.5$. Same parameters as in Fig.~\ref{FigEf}.}\label{FigPFA}\end{figure}

To be more quantitative, we give here the precise definition of the PFA free energy at a distance $d$ in the sphere-grating configuration
\begin{equation}\label{PFA1}{\cal F}^{(1)}_\text{PFA}(d)=2\pi R_S\int_d^{d+R_S}{\cal E}_{PG}(z)dz,\end{equation}
where ${\cal E}_{PG}(z)$ represents the exact Casimir free energy per unit area between a gold plane and the grating under scrutiny in our calculation, placed at distance $z$ from each other. One {\color{black} can} perform an even rougher approximation, by also {\color{black} replacing} the grating
by a set of planar half spaces,
one at distance $d$ and one at distance $d+h$ from the sphere {\color{black}(see Fig.~\ref{Geometry})}, and averaging the results using the filling factor $f$ as a weight. In
{\color{black} this double proximity force approximation, the Casimir energy reads}
\begin{equation}\begin{split}\label{PFA2}&{\cal F}^{(2)}_\text{PFA}(d)\\
&=2\pi R_S\Bigl[f\int_d^{d+R_S}{\cal E}_{PP}(z)dz\\
&\,+(1-f)\int_{d+h}^{d+h+R_S}{\cal E}_{PP}(z)dz\Bigr]\\
&=2\pi R_S\Bigl[f\Bigl(D_{PP}(d)-D_{PP}(d+R_S)\Bigr)\\
&\,+(1-f)\Bigl(D_{PP}(d+h)-D_{PP}(d+h+R_S)\Bigr)\Bigr].\end{split}\end{equation}
In this expression we have defined the function $D_{PP}(z)$ {\color{black} by the condition} $-\frac{\partial D_{PP}(z)}{\partial z}={\cal E}_{PP}(z)$. While this function can be expressed analytically in the case of the plane-plane interaction, this is no longer true in the case of the plane-grating geometry. Thus, in order to {\color{black} compute the single-PFA expression ${\cal F}^{(1)}_\text{PFA}$ from \eqref{PFA1},} one needs to calculate the energy ${\cal E}_{PG}(z)$
for several distances and then integrate it numerically from $d$ to $d+R_S$.

The comparison between the exact results and the two PFA approaches \eqref{PFA1} and \eqref{PFA2} {\color{black}is} presented in Fig.~\ref{FigPFA}, where we plot the ratio
\begin{equation}\label{etadef}
\eta=\frac{{\cal F}(d)}{{\cal F}_\text{PFA}(d)}
\end{equation}
as function of distance. We calculate for the same filling factors used in Fig.~\ref{FigEf}, including $f=1$ corresponding to a plane-sphere configuration. The figure shows that the PFA overestimates the interaction energy and becomes more accurate as $d/R_S$ decreases, as expected. More interestingly, lower values of the filling factor prove to give lower values for the ratio $\eta$ when taking the single PFA \eqref{PFA1}. In other words, the introduction of a more complex geometry, namely the structuring of the plane into a grating, makes the replacement of the exact spherical surface by a set of parallel planes less accurate for the description of the interaction. For example, for the closest distance $d=1\,\mu$m (corresponding to $d/R_S=0.2$) shown in Fig.~\ref{FigPFA}, we have a ratio $\eta_E=0.89$ for the plane-sphere case ($f=1$) which drops to $\eta_E=0.84$ in the case of $f=0.3$. At this point, it is important to note, in connection with a recent sphere-grating experiment ~\cite{IntravaiaNatComm14}, that the theoretical approach based on the single PFA \eqref{PFA1} leads to a huge overestimation of the experimental data for the Casimir force gradient. In this experiment, the interaction was probed for distances such that $d/R_S < 10^{-2}.$ In this range, it is known that PFA is sufficiently accurate when describing the plane-sphere geometry \cite{DeccaPRL07}, but no indication about the accuracy of \eqref{PFA1} for the sphere-grating configuration was available so far. Although our numerical results only cover distance-to-radius ratios much larger than those probed in the experiment, they indicate that PFA is indeed worse when a plane surface is replaced by a grating, even when the diffraction by the grating is fully taken into account. However, the discrepancy we have found is most likely too small to explain the theory-experiment disagreement found in \cite{IntravaiaNatComm14}.

In the inset of Figure \ref{FigPFA}, we compare the values of $\eta$ when taking either the single \eqref{PFA1} or double \eqref{PFA2} PFA result in the denominator of \eqref{etadef} (green triangles and orange circles, respectively). We also plot $\eta$ for the standard plane-sphere geometry (grey squares). Note that the double PFA is surprisingly more accurate than the single PFA. This is clearly a consequence of the fact that the PFA approach underestimates the interaction energy in the plane-grating geometry \cite{Davids2010}. In fact, when both the sphere and the grating are approximated by planar surfaces, the corresponding errors have opposite signs and tend to compensate each other.

\subsection{Lateral Casimir force}

As already highlighted before, the presence of the grating breaks the translational symmetry along the $x$ axis. Since the Casimir energy depends on the sphere lateral position $x_S,$ the Casimir force acquires a lateral component along the $x$ axis given by
\begin{equation}F_x(x_S,d)=-\frac{\partial\mathcal{F}(x_S,d)}{\partial x_S}.\end{equation}
For very large spheres close to the validity regime of the PFA, the lateral dependence is averaged out by the very large nearly-plane spherical surface. On the other hand, since our approach fully accounts for the curvature of the sphere and allows to consider different sphere radii, we can analyze the lateral dependence of the Casimir energy in a variety of configurations leading to possibly measurable lateral forces. In the limit of very small radii, the sphere plays the role of a local probe of the field fluctuations. By considering finite-size metallic spheres, we expect to find lateral forces larger than in the atomic case \cite{BuhmannarXiv15}, thus improving the prospect for a possible experiment (note that the lateral Casimir force between two metallic gratings has been measured \cite{Chen2002}).

\begin{figure}[h]\centering
\includegraphics[height=7.5cm]{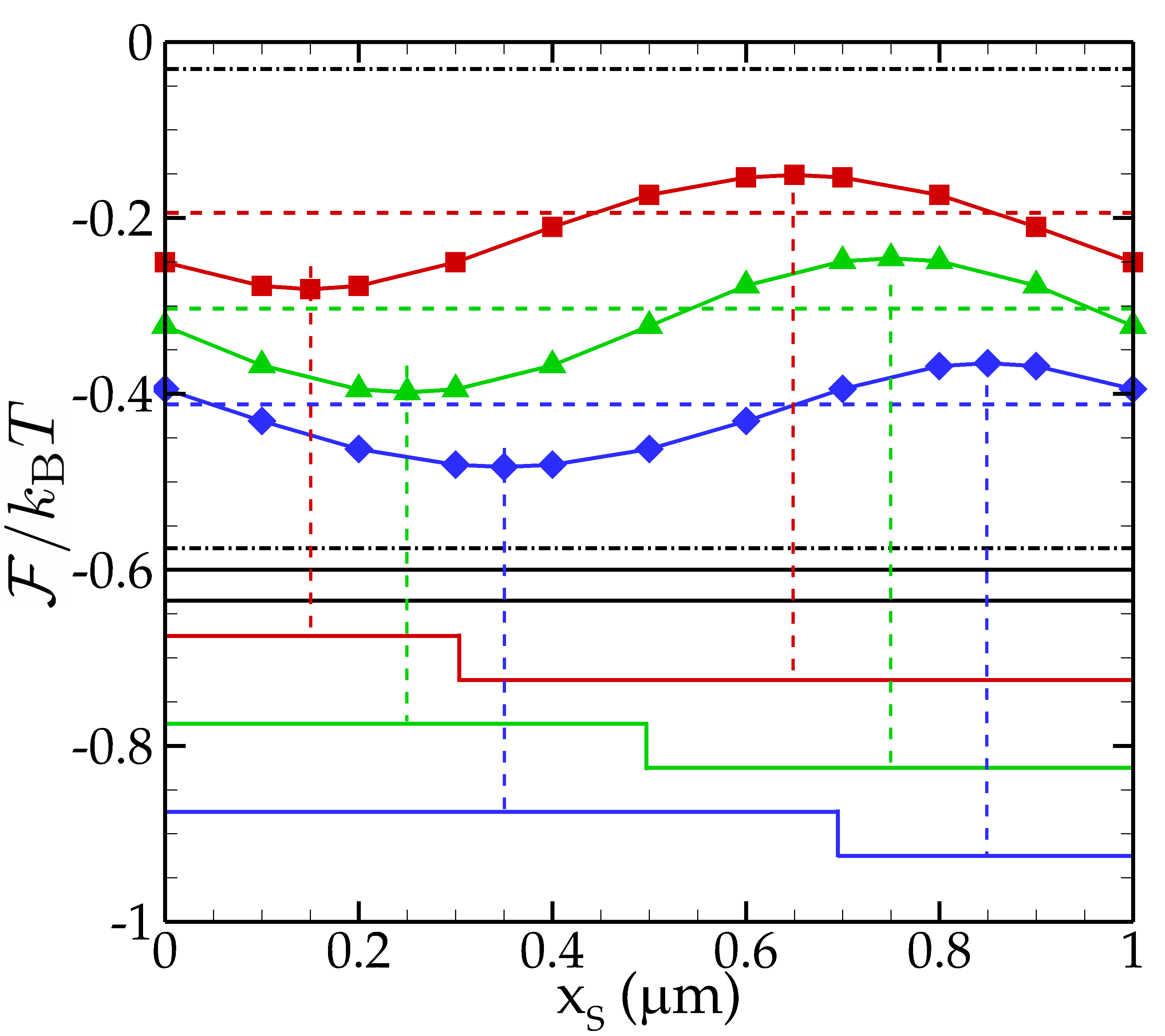}
\caption{Dependence of the Casimir energy on the sphere center lateral position $x_S$. The sphere radius and distance are $R_S=500\,$nm and $d=200\,$nm. The grating parameters are $D=1\,\mu$m, $h=500\,$nm and $f=0.3$ (red squares), 0.5 (green triangles), 0.7 (blue diamonds). The exact results are compared to several plane-sphere energies (see main text for details).}\label{FigShift}\end{figure}

Let us consider a sphere of radius $R_S=500\,$nm placed at a distance $d=200\,$nm from a grating having period $D=1\,\mu$m and depth $h=500\,$nm. We consider three different filling factors $f=0.3,0.5,0.7$ and plot the Casimir energy as a function of $x_S$ over one period in Fig.~\ref{FigShift}. As expected, we observe indeed a dependence on $x_S$ and we confirm the periodicity of $\cal F$. Moreover, we clearly see that the minima and maxima of the energy for the three filling factors are clearly associated to the center of the two grating regions. In particular, the energy is minimized when the center of the sphere is located at the center of the higher region of the grating (which is then a stable equilibrium position with respect to displacements along the $x$ axis), while it is maximized when the sphere is located at the center of the lower region (unstable equilibrium position). In Fig.~\ref{FigShift}, we also show reference energy values obtained for plane-sphere configurations. The lower and higher values (dot-dashed horizontal lines) correspond to the sphere-plane energies at the two distances $d$ and $d+h$. Moreover, the dashed value around which each curve is oscillating is the weighted average of these two sphere-plane results using the appropriate filling factors. We see that the two sphene-plane results make it possible to predict very roughly the region of oscillation, and the weighted averages give a good estimate of the center of oscillation.

\begin{figure}[h]\centering
\includegraphics[height=7.5cm]{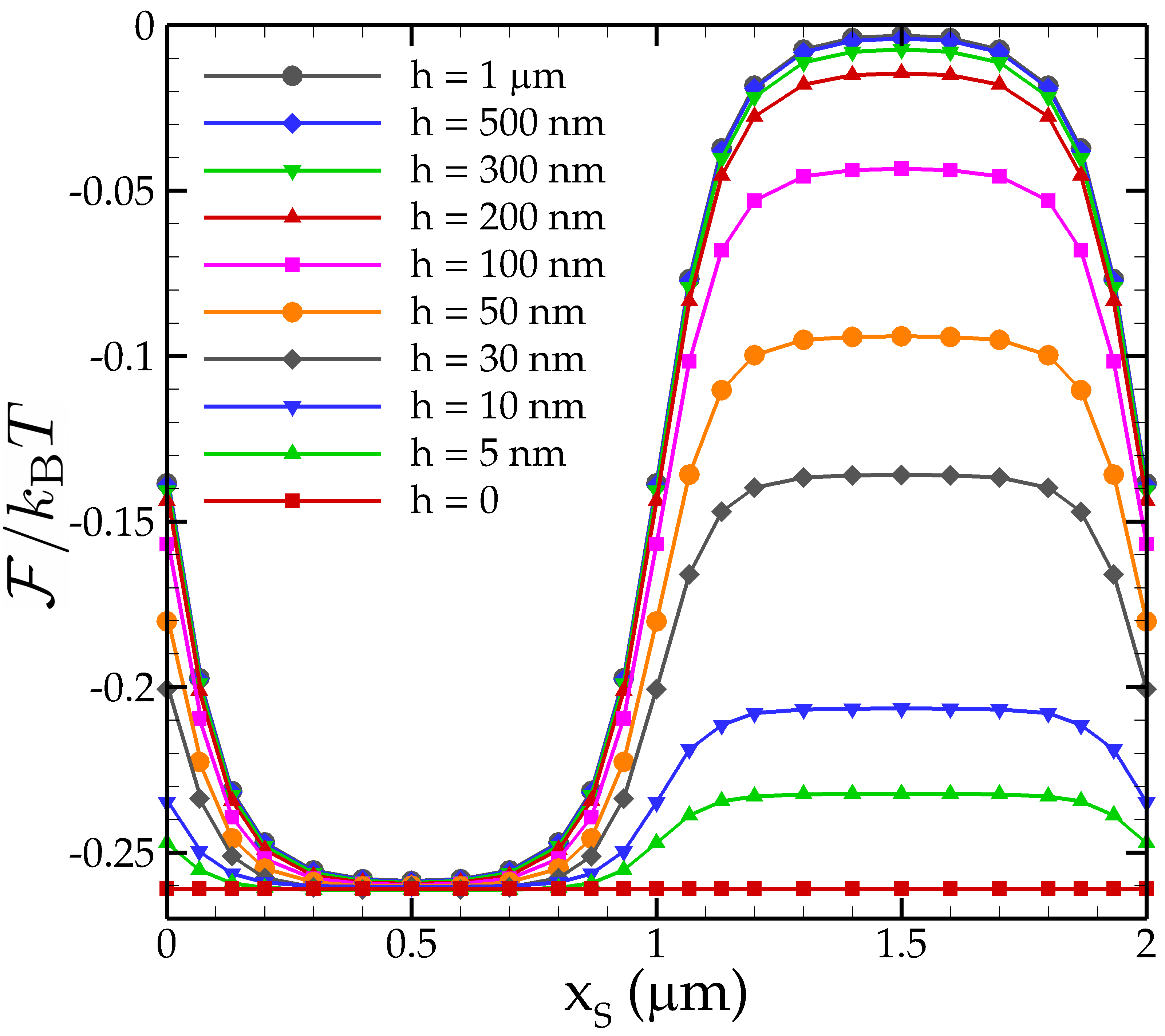}
\caption{Casimir energy variation with the sphere lateral position for a nanosphere of radius $R_S=100\,$nm at a distance $d=100\,$nm from a grating of period $D=2\,\mu$m and filling factor $f=0.5$. The different curves correspond to different grating heights (see legend).}\label{FigPlateau}\end{figure}

\subsection{Nanosphere as a local probe of field fluctuations}

We now consider the case in which both the radius of the sphere and the sphere-grating distance are small compared to the grating period. In particular, we take into account a nanosphere of radius $R_S=100\,$nm placed at a distance of $d=100\,$nm from a grating of period $D=2\,\mu$m and $f=0.5$. Since the ratios $D/R_S$ and $D/d$ are equal to 20, we expect the sphere-grating energy to show two flat parts around the center of the two regions of the grating, with two transition regions close to the corners. We have studied this behavior for different values of the grating height $h$. The results are presented in Fig.~\ref{FigPlateau}.

\begin{figure}[h]\centering
\includegraphics[height=7.5cm]{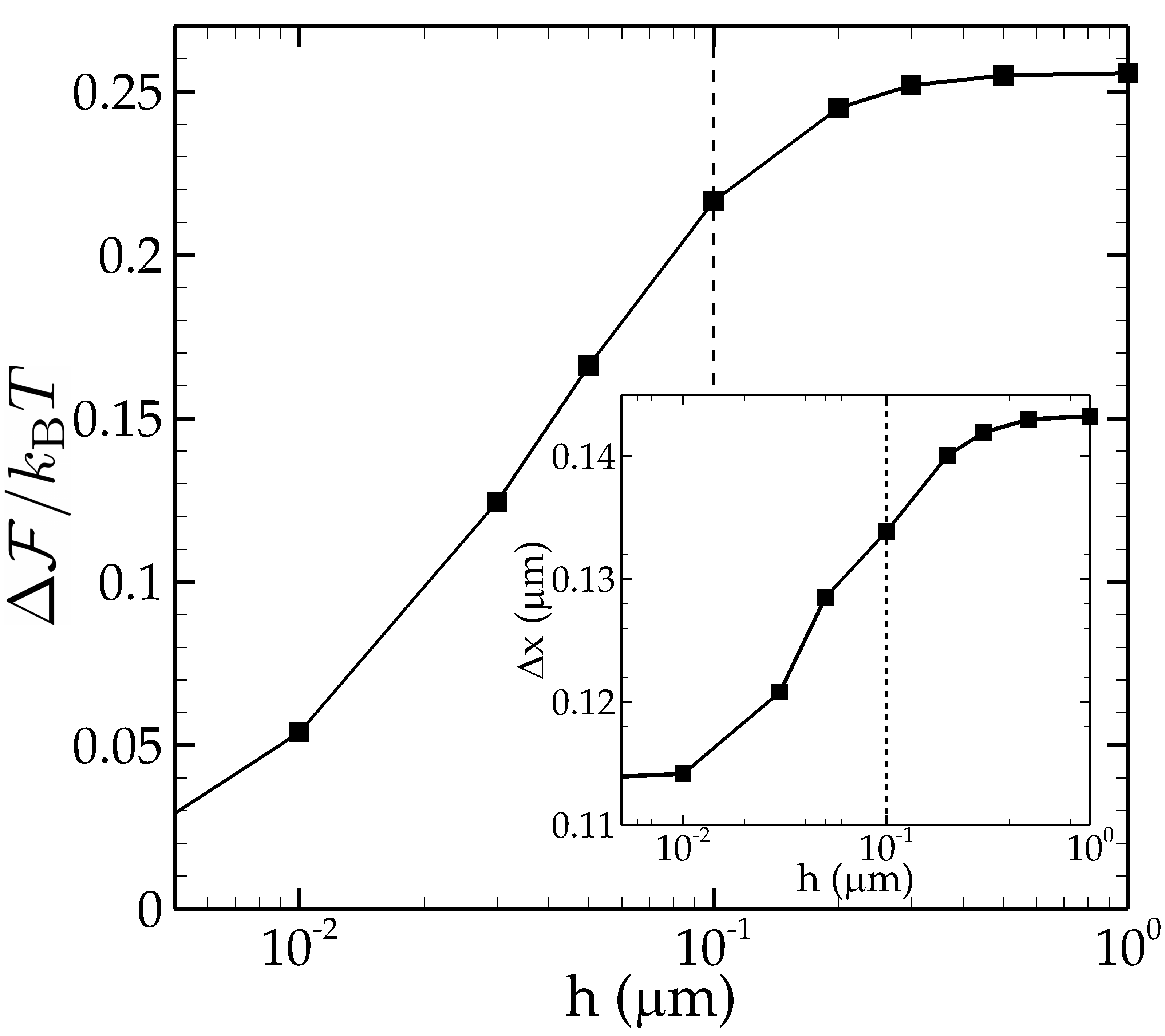}
\caption{In the main part, oscillation amplitude \eqref{DeltaE} of the Casimir energy represented in Fig.~\ref{FigPlateau} as a function of $h$. In the inset, transition length \eqref{TransL} as a function of $h$.}\label{FigDeltaE}\end{figure}

First of all, we clearly observe that in the case of $h=0$ we have no dependence on $x_S$, since this case corresponds to a plane-sphere configuration. On the other hand, we see indeed a dependence on the lateral {\color{black}position} for all the other values of $h$, which is already visible at values as low as $h=5\,$nm. Then, the curves are strongly modified by increasing the value of the height, and clearly seem to reach an asymptotic behavior for $h$ of the order of 500\,nm. Such asymptotic behavior suggests that the bottom of the grating is not affecting the interaction energy of a nanosphere sitting on top of the center of the plateau region, as long as $h$ is sufficiently large.

Moreover, we observe the existence of two regions with a slow variation with $x_S$, close to the centers of the high and low part of the grating. As expected, the width of this nearly flat region decreases with $h$, whereas the difference between the low and high values of the energy increases. In order to make this analysis quantitative, we have considered the difference between the extreme values of the energy as a function of $h$. This difference is clearly defined as
\begin{equation}\label{DeltaE}\Delta\mathcal{F}=\mathcal{F}(1.5\,\mu\text{m})-\mathcal{F}(0.5\,\mu\text{m}),\end{equation}
and is plotted in Fig.~\ref{FigDeltaE}. This difference is exactly zero for $h=0$ (not visible in logarithmic scale) and goes to an asymptotic value reached around 500\,nm as observed before from Fig.~\ref{FigPlateau}. We have also studied the characteristic transition length between the two regions:
\begin{equation}\label{TransL}
\Delta x=x_2-x_1,
\end{equation}
where $x_1$ and $x_2$ are defined by
\begin{equation}
\begin{split}
\mathcal{F}(x_1)&=\frac{\mathcal{F}(0.5\,\mu\text{m})+\mathcal{F}(1.0\,\mu\text{m})}{2}\\
\mathcal{F}(x_2)&=\frac{\mathcal{F}(1.0\,\mu\text{m})+\mathcal{F}(1.5\,\mu\text{m})}{2}.\end{split}\end{equation}
In other words, $x_2$ ($x_1$) is the point at which the energy, starting from $x=1\,\mu$m and moving to larger (smaller) values of $x_S$, has increased (decreased) by half of the value it would get to arrive at its maximum (minimum). We see the inset of Fig.~\ref{FigDeltaE} that this quantity has a behavior similar to the one of $\Delta\mathcal{F}$. It is interesting to remark that its asymptotic value is around 140\,nm: it is then of the order of the sphere geometric parameters $R_S$ and $d$, confirming the behavior of the sphere as a local probe of the field.

\begin{figure}[h]\centering
\includegraphics[height=7.5cm]{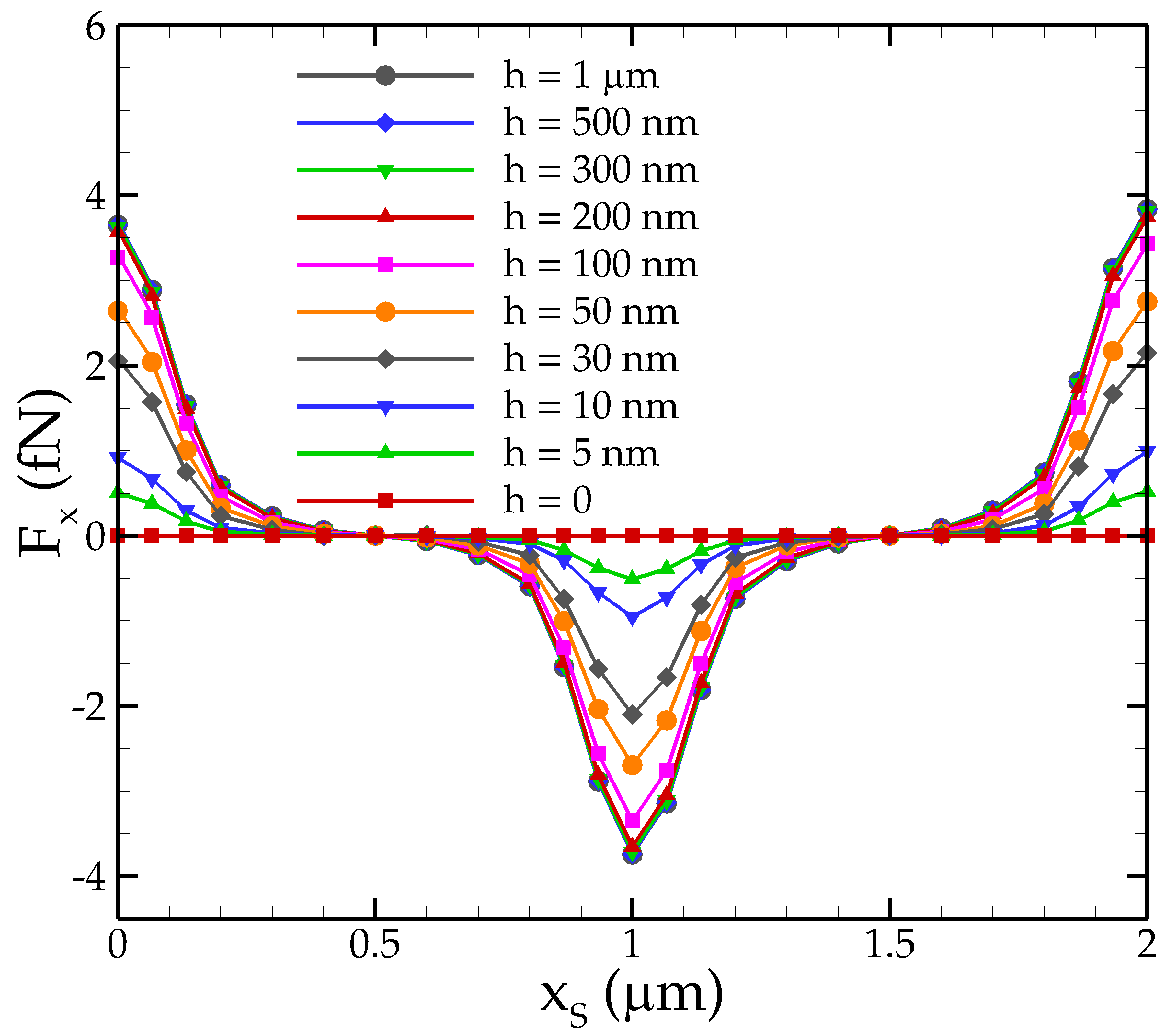}
\caption{Lateral Casimir force for a nanosphere of radius $R_S=100\,$nm at a distance $d=100\,$nm from a grating of period $D=2\,\mu$m and filling factor $f=0.5$. The different curves correspond to different grating heights (see legend).}\label{FigPlateauf}\end{figure}

Finally, in Fig.~\ref{FigPlateauf} we show the lateral force acting on the sphere, computed from the numerical derivative of the curves for the energy shown in Fig.~\ref{FigPlateau}. The force points towards the center of the plateau regions and its magnitude is maximum on top of the corners. For the nanospheres considered in Figs.~\ref{FigPlateauf} and \ref{FigPlateau}, the maximum lateral force is close to 4\,fN, hence considerably smaller than the values measured in \cite{Chen2002} for the more complex geometry involving a second grating imprinted on the surface of a very large sphere. Nevertheless, it is remarkable that a lateral force exists even with a single grating, as long as the sphere surface is sufficiently small and does not average out the grating profile. One can still expect a considerable increase in the value of the force by replacing our dielectric grating by a metallic one (as in \cite{Chen2002,Chiu2010}) and by optimizing the sphere radius and the grating period and filling factor.

\section{Conclusions}\label{SecConcl}

We derived the exact sphere-grating Casimir interaction energy, which we numerically investigated for a metallic sphere and a dielectric grating.
We considered separation/radius aspect ratios $d/R_S\stackrel{>}{\scriptscriptstyle\sim} 0.2$. These values of $d/R_S$ are still more than one order of magnitude larger than the aspect ratios probed in existing Casimir force experiments \cite{IntravaiaNatComm14,Chen2002,Chiu2010,Chan2008,Chan2010,Banishev2013}. Nonetheless, our results show that replacing the plane surface by a nanostructured grating degrades the quality of the PFA description of the sphere curvature.

The sphere is shown to be a probe of several geometric effects in Casimir interaction. In particular, its finite size has a startling consequence: a lateral Casimir force appears as long as the sphere surface does not average out the grating profile.

This work opens to several interesting developments, for instance by considering a metallic grating in order to increase both the normal and the lateral Casimir interaction.

\begin{acknowledgments}
The authors acknowledge financial support from the Julian Schwinger Foundation. P.A.M.N. thanks the group ``Theory of Light-Matter and Quantum Phenomena'' of the Laboratoire Charles Coulomb for hospitality during his stay in Montpellier, and CNPq and CNE-FAPERJ for financial support.
\end{acknowledgments}

\appendix*

\section{}\label{AppSphere}

We provide here all the details concerning the choice of the plane-wave and spherical bases, as well as the detailed derivation of the projection coefficients between the two representations.

\subsection{Plane-wave basis}

We start providing the explicit expression of the electric field, which we first decompose with respect to frequency, working only with positive frequencies
\begin{equation}\mathbf{E}(\mathbf{R},t)=2\Rea\Biggl[\int_0^{+\infty}\frac{d\omega}{2\pi}\exp(-i\omega t)\mathbf{E}(\mathbf{R},\omega)\Biggr].\end{equation}
The single-frequency component $\mathbf{E}(\mathbf{R},\omega)$ is then decomposed with respect to the parallel wavevector $\mathbf{k}$, the direction of propagation $\phi$ and the polarization $p$
\begin{equation}\label{DefE}\mathbf{E}(\mathbf{R},\omega)=\sum_{\phi,p}\int\frac{d^2\mathbf{k}}{(2\pi)^2}
\exp(i\mathbf{K}^\phi\cdot\mathbf{R})\hat{\bbm[\epsilon]}_p^\phi(\mathbf{k},\omega)E_p^\phi(\mathbf{k},\omega).\end{equation}
As a general rule, the sum on $\phi$ runs over the values
$\{+,-\}$, the sum on $p$ over the values $\{1,2\}$. For the polarization vectors
$\hat{\bbm[\epsilon]}_p^\phi(\mathbf{k},\omega)$ appearing in Eq.~\eqref{DefE} we adopt the following definitions
\begin{equation}\label{PolVect}\begin{split}\hat{\bbm[\epsilon]}_\TE^\phi(\mathbf{k},\omega)&=\hat{\mathbf{z}}\times\hat{\mathbf{k}}=\frac{1}{k}(-k_y\hat{\mathbf{x}}+k_x\hat{\mathbf{y}}),\\
\hat{\bbm[\epsilon]}_\TM^\phi(\mathbf{k},\omega)&=\frac{1}{K}\hat{\bbm[\epsilon]}_\TE^\phi(\mathbf{k},\omega)\times\mathbf{K}^\phi=\frac{1}{K}(-k\hat{\mathbf{z}}+\phi
k_z\hat{\mathbf{k}}),\end{split}\end{equation} where
$\hat{\mathbf{x}}$, $\hat{\mathbf{y}}$ and $\hat{\mathbf{z}}$ are
the unit vectors along the directions $x$, $y$ and $z$
respectively and $\hat{\mathbf{k}}=\mathbf{k}/k$. The following relation holds
\begin{equation}\label{CurlEps}\hat{\mathbf{K}}^\phi\times\hat{\bbm[\epsilon]}_p^\phi(\mathbf{k},\omega)=\hat{\bbm[\beta]}_p^\phi(\mathbf{k},\omega)=(-1)^p\hat{\bbm[\epsilon]}_{S(p)}^\phi(\mathbf{k},\omega),\end{equation}
being $S(p)$ the function which switches between the two
polarization, acting as $S(1)=2$ and $S(2)=1$.

The expression of the single-frequency component of the magnetic field can be easily deduced from Maxwell's equations. It reads
\begin{equation}\label{DefB}\begin{split}\mathbf{B}(\mathbf{R},\omega)=\frac{\sqrt{\varepsilon(\omega)}}{c}\sum_{\phi,p}&\int\frac{d^2\mathbf{k}}{(2\pi)^2}
\exp(i\mathbf{K}^\phi\cdot\mathbf{R})\\
&\,\times\hat{\bbm[\beta]}_p^\phi(\mathbf{k},\omega)E_p^\phi(\mathbf{k},\omega).\end{split}\end{equation}

\subsection{Spherical-wave basis}

The presence of the sphere makes it natural to introduce a spherical-wave basis. In this basis a mode is identified by the set $(\omega,\ell,m,P,s)$, where $\ell$ and $m$ are as usually associated to the eigenvalues of the angular-momentum operators $\mathbf{L}^2$ and $L_z$, while $P$ denotes the spherical polarization, taking the values $P=\text{E}$ (electric) and $P=\text{M}$ (magnetic). Finally, as stated in the main text, $s$ is associated to regular ($s=$\,reg) or outgoing ($s=$\,out) modes. In this mode decomposition, we write the regular or outgoing electric field as
\begin{equation}\mathbf{E}^{(s)}(\mathbf{R},\omega)=\sum_{\ell=1}^{+\infty}\sum_{m=-\ell}^\ell\sum_{P\in\{\text{E},\text{M}\}}E^{(s)}_{\ell mP}(\omega)\mathbf{F}^{(s)}_{\ell mP}(\mathbf{R}).\end{equation}
The mode functions $\mathbf{F}^{(s)}_{\ell mP}$ are defined as follows
\begin{equation}\label{MN}\begin{split}\mathbf{F}^{(s)}_{\ell m\text{M}}(\mathbf{R})&=\nabla\times\bigl(\mathbf{R}\psi^{(s)}_{\ell m}(\mathbf{R})\bigr),\\
\mathbf{F}^{(s)}_{\ell m\text{E}}(\mathbf{R})&=\frac{1}{K}\nabla\times\mathbf{F}^{(s)}_{\ell m\text{M}}(\mathbf{R}).\end{split}\end{equation}
The scalar function $\psi^{(s)}_{\ell m}$ appearing in these equations is a solution of Helmoltz equation and can be cast in the form
\begin{equation}\psi^{(s)}_{\ell m}(\mathbf{R})=\frac{i^\ell}{\sqrt{\ell(\ell+1)}}z^{(s)}_\ell(KR)Y_{\ell m}(\theta,\varphi),\end{equation}
where for regular modes $z^{(\text{reg})}_\ell=j_\ell$, while for outgoing modes (diverging at the origin) we have $z^{(\text{out})}_\ell=h_\ell^{(1)}$. In this expression we are working in spherical coordinates. The modes \eqref{MN} can be calculated explicitly and they read
\begin{equation}\label{MN2}\begin{split}\mathbf{F}^{(s)}_{\ell m\text{M}}(\mathbf{R})&=\frac{i^\ell}{\sqrt{\ell(\ell+1)}}\Bigl[\frac{im}{\sin\theta}z^{(s)}_\ell(KR)Y_{\ell m}(\theta,\varphi)\hat{\bbm[\theta]}\\
&\,-z^{(s)}_\ell(KR)\frac{\partial Y_{\ell m}(\theta,\varphi)}{\partial\theta}\hat{\bbm[\varphi]}\Bigr],\\
\mathbf{F}^{(s)}_{\ell m\text{E}}(\mathbf{R})&=\frac{i^\ell}{\sqrt{\ell(\ell+1)}}\Bigl[\frac{1}{KR}z^{(s)}_\ell(KR)\ell(\ell+1)Y_{\ell m}(\theta,\varphi)\hat{\mathbf{R}}\\
&\,+\frac{1}{KR}\frac{d}{dR}\bigl(R\,z^{(s)}_\ell(KR)\bigr)\frac{\partial Y_{\ell m}(\theta,\varphi)}{\partial\theta}\hat{\bbm[\theta]}\\
&\,+\frac{im}{\sin\theta}\frac{1}{KR}\frac{d}{dR}\bigl(R\,z^{(s)}_\ell(KR)\bigr)Y_{\ell m}(\theta,\varphi)\hat{\bbm[\varphi]}\Bigr].\end{split}\end{equation}
It can be shown that these functions satisfy the following conditions (where $S(\text{E})=\text{M}$ and $S(\text{M})=\text{E}$)
\begin{equation}\label{CurlF}\nabla\times\mathbf{F}^{(s)}_{\ell mP}(\mathbf{R})=K\mathbf{F}^{(s)}_{\ell mS(P)}(\mathbf{R}),\end{equation}
and the orthogonality and normalization properties
\begin{equation}\begin{split}
\int_0^\pi d\theta\sin\theta&\int_0^{2\pi}d\varphi\,\mathbf{F}^{(s)*}_{\ell m\text{E}}(\mathbf{R})\cdot\mathbf{F}^{(s)}_{\ell'm'\text{M}}(\mathbf{R})=0,\\
\int_0^\pi d\theta\sin\theta&\int_0^{2\pi}d\varphi\,\mathbf{F}^{(s)*}_{\ell m\text{M}}(\mathbf{R})\cdot\mathbf{F}^{(s)}_{\ell'm'\text{M}}(\mathbf{R})=\\
&=\delta_{\ell,\ell'}\delta_{m,m'}\bigl(z^{(s)}_\ell(KR)\bigr)^2,\\
\int_0^\pi d\theta\sin\theta&\int_0^{2\pi}d\varphi\,\mathbf{F}^{(s)*}_{\ell m\text{E}}(\mathbf{R})\cdot\mathbf{F}^{(s)}_{\ell'm'\text{E}}(\mathbf{R})=\\
&=\delta_{\ell,\ell'}\delta_{m,m'}\frac{1}{K^2R^2}\Bigl[\Bigl(\frac{d}{dR}\bigl(R\,z^{(s)}_\ell(KR)\bigr)\Bigr)^2\\
&\,+\ell(\ell+1)\bigl(z^{(s)}_\ell(KR)\bigr)^2\Bigr].
\end{split}\end{equation}
As a consequence, the amplitude $E^{(s)}_{\ell mP}$ of a given field $\mathbf{E}^{(s)}(\mathbf{R},\omega)$ with respect to a mode function $\mathbf{F}^{(s)}_{\ell mP}$ can be calculated as
\begin{equation}E^{(s)}_{\ell mP}=\frac{\int_0^\pi d\theta\sin\theta\int_0^{2\pi}d\varphi\,\mathbf{F}^{(s)*}_{\ell mP}(\mathbf{R})\cdot\mathbf{E}^{(s)}(\mathbf{R},\omega)}{\int_0^\pi d\theta\sin\theta\int_0^{2\pi}d\varphi\,|\mathbf{F}^{(s)}_{\ell mP}(\mathbf{R})|^2}.\end{equation}

\subsection{Projection of a plane wave on spherical waves}

Because of the mixture of symmetries typical of our physical system, we have to face the problem of calculating the matrix elements describing the projections of any basis function of one set (plane or spherical waves) on the functions of the other set. We start here from the problem of decomposing a plane wave of frequency $\omega$, wavevector $\mathbf{K}^\phi$ and polarization $p$ on the set of spherical waves. Since a plane wave is regular at the origin, this decomposition will involve only spherical modes having $s=$\,reg. More explicitly we want to derive the coefficients defined by the following decomposition:
\begin{equation}\label{Decomp1}\exp(i\mathbf{K}^\phi\cdot\mathbf{R})\hat{\bbm[\epsilon]}_p^\phi(\mathbf{k},\omega)=\sum_{\ell,m,P}\bra{\ell,m,P,\text{reg}}\mathbf{k},p,\phi\rangle\mathbf{F}^{(\text{reg})}_{\ell mP}(\mathbf{R}).\end{equation}

To this aim, we start by expressing the plane-wave unit vectors with respect to spherical unit vectors. This gives
\begin{equation}\label{UnVect}\begin{split}\hat{\bbm[\epsilon]}_\TE^\phi(\mathbf{k},\omega)&=\sin\theta\sin(\varphi-\varphi_\mathbf{k})\hat{\mathbf{R}}+\cos\theta\sin(\varphi-\varphi_\mathbf{k})\hat{\bbm[\theta]}\\
&\,+\cos(\varphi-\varphi_{\mathbf{k}})\hat{\bbm[\varphi]},\\
\hat{\bbm[\epsilon]}_\TM^\phi(\mathbf{k},\omega)&=\Bigl(-\sin\theta^\phi_\mathbf{k}\cos\theta+\cos\theta^\phi_\mathbf{k}\sin\theta\cos(\varphi-\varphi_\mathbf{k})\Bigr)\hat{\mathbf{R}}\\
&\,+\Bigl(\sin\theta^\phi_\mathbf{k}\sin\theta+\cos\theta^\phi_\mathbf{k}\cos\theta\cos(\varphi-\varphi_\mathbf{k})\Bigr)\hat{\bbm[\theta]}\\
&\,-\cos\theta^\phi_\mathbf{k}\sin(\varphi-\varphi_\mathbf{k})\hat{\bbm[\varphi]}.\end{split}\end{equation}
In these expressions we have introduced the angles $\theta_\mathbf{k}$ and $\varphi_\mathbf{k}$, defined by the relation
\begin{equation}\label{Angles}\begin{split}&\mathbf{K}^\phi=K(\sin\theta^\phi_\mathbf{k}\cos\varphi_\mathbf{k},\sin\theta^\phi_\mathbf{k}\sin\varphi_\mathbf{k},\cos\theta^\phi_\mathbf{k})\\
&\Rightarrow\quad\sin\theta^\phi_\mathbf{k}=\frac{k}{K},\quad\cos\theta^\phi_\mathbf{k}=\frac{\phi k_z}{K},\\\
&\hspace{.8cm}\cos\varphi_\mathbf{k}=\frac{k_x}{k},\quad\sin\varphi_\mathbf{k}=\frac{k_y}{k}.\end{split}\end{equation}
If we move to an imaginary frequency $\omega=i\xi$ the angle $\theta^\phi$ is defined as
\begin{equation}\sin\theta^\phi_\mathbf{k}=-i\frac{ck}{\xi},\quad\cos\theta^\phi_\mathbf{k}=\frac{\phi c\kappa}{\xi},\quad k_z=i\kappa=i\sqrt{\frac{\xi^2}{c^2}+k^2}.\end{equation}
For an arbitrary frequency we deduce
\begin{equation}\mathbf{K}^\phi\cdot\mathbf{R}=KR\Bigl(\sin\theta^\phi_\mathbf{k}\sin\theta\cos(\varphi-\varphi_\mathbf{k})+\cos\theta^\phi_\mathbf{k}\cos\theta\Bigr).\end{equation}
In order to calculate the matrix elements defined in Eq.~\eqref{Decomp1} we will make use of the following result
\begin{equation}\label{Intphi}\begin{split}&\int_0^{2\pi}d\varphi\,e^{i\beta\cos(\varphi-\varphi_\mathbf{k})}e^{-im\varphi}\begin{pmatrix}\sin(\varphi-\varphi_\mathbf{k})\\\cos(\varphi-\varphi_\mathbf{k})\\1\end{pmatrix}\\
&=-\pi i^me^{-im\varphi_\mathbf{k}}\begin{pmatrix}\bigl(J_{m-1}(\beta)+J_{m+1}(\beta)\bigr)\\i\bigl(J_{m-1}(\beta)-J_{m+1}(\beta)\bigr)\\-2J_m(\beta)\end{pmatrix},\end{split}\end{equation}
$J_n(x)$ being the ordinary Bessel function of index $n$. Taking the curl of both sides of Eq.~\eqref{Decomp1} and using Eqs.~\eqref{CurlEps} and \eqref{CurlF} we have easily
\begin{equation}\begin{split}\exp(i\mathbf{K}^\phi\cdot\mathbf{R})&\hat{\bbm[\epsilon]}_p^\phi(\mathbf{k},\omega)=\sum_{\ell,m,P}i(-1)^p\\
&\times\bra{\ell,m,S(P),\text{reg}}\mathbf{k},S(p),\phi\rangle\mathbf{F}^{(\text{reg})}_{\ell mP}(\mathbf{R}),\end{split}\end{equation}
from which we deduce the property
\begin{equation}\label{PropProj}\bra{\ell,m,P,\text{reg}}\mathbf{k},p,\phi\rangle=i(-1)^p\bra{\ell,m,S(P),\text{reg}}\mathbf{k},S(p),\phi\rangle.\end{equation}

\begin{widetext}
Let us start with the calculation of $\bra{\ell,m,\text{M},\text{reg}}\mathbf{k},\TE,\phi\rangle$.We have
\begin{equation}\begin{split}&\exp(i\mathbf{K}^\phi\cdot\mathbf{R})\mathbf{F}^{(\text{reg})*}_{\ell m\text{M}}(\mathbf{R})\cdot\hat{\bbm[\epsilon]}_\TE^\phi(\mathbf{k},\omega)=\exp[iKR\sin\theta^\phi_\mathbf{k}\sin\theta\cos(\varphi-\varphi_\mathbf{k})]\exp[iKR\cos\theta^\phi_\mathbf{k}\cos\theta]\\
&\quad\times\frac{i^{-\ell}}{\sqrt{\ell(\ell+1)}}\Bigl(-\frac{im
\cos\theta}{\sin\theta}\sin(\varphi-\varphi_\mathbf{k})Y_{\ell m}(\theta,0)-\cos(\varphi-\varphi_{\mathbf{k}})\frac{\partial Y_{\ell m}(\theta,0)}{\partial\theta}\Bigr)j_\ell(KR)e^{-im\varphi},\end{split}\end{equation}
We now use Eq.~\eqref{Intphi} and the two following relations
\begin{equation}\label{RelY}-\frac{2m\cos\theta}{\sin\theta}Y_{\ell m}(\theta,\varphi)=e^{-i\varphi}\sqrt{(\ell-m)(\ell+m+1)}Y_{\ell,m+1}(\theta,\varphi)+e^{i\varphi}\sqrt{(\ell+m)(\ell-m+1)}Y_{\ell,m-1}(\theta,\varphi),\end{equation}
\begin{equation}\label{dY}\begin{split}\frac{\partial Y_{\ell m}(\theta,\varphi)}{\partial\theta}&=\frac{m
\cos\theta}{\sin\theta}Y_{\ell m}(\theta,\varphi)+e^{-i\varphi}\sqrt{(\ell-m)(\ell+m+1)}Y_{\ell,m+1}(\theta,\varphi)\\
&=\frac{1}{2}\Bigl(e^{-i\varphi}\sqrt{(\ell-m)(\ell+m+1)}Y_{\ell,m+1}(\theta,\varphi)-e^{i\varphi}\sqrt{(\ell+m)(\ell-m+1)}Y_{\ell,m-1}(\theta,\varphi)\Bigr),\end{split}\end{equation}
to get
\begin{equation}\begin{split}&\int_0^{2\pi}d\varphi\,\exp(i\mathbf{K}^\phi\cdot\mathbf{R})\mathbf{F}^{(\text{reg})*}_{\ell m\text{M}}(\mathbf{R})\cdot\hat{\bbm[\epsilon]}_\TE^\phi(\mathbf{k},\omega)=-\pi i^{m+1}e^{-im\varphi_\mathbf{k}}\exp[iKR\cos\theta^\phi_\mathbf{k}\cos\theta]j_\ell(KR)\\
&\quad\times\frac{i^{-\ell}}{\sqrt{\ell(\ell+1)}}\Bigl[\sqrt{(\ell+m)(\ell-m+1)}J_{m-1}(\beta)Y_{\ell,m-1}(\theta,0)+\sqrt{(\ell-m)(\ell+m+1)}J_{m+1}(\beta)Y_{\ell,m+1}(\theta,0)\Bigr].\end{split}\end{equation}
Using the result \cite{NevesJPhysA06}
\begin{equation}\label{Intej}\int_0^\pi d\theta\,\sin\theta\,\exp[ia\cos\alpha\cos\theta]Y_{\ell m}(\theta,\varphi)J_m(a\sin\alpha\sin\theta)=2i^{\ell-m}Y_{\ell m}(\alpha,\varphi)j_\ell(a),\end{equation}
we conclude, using again Eq.~\eqref{RelY},
\begin{equation}\begin{split}\bra{\ell,m,\text{M},\text{reg}}\mathbf{k},\TE,\phi\rangle&=\frac{2\pi}{\sqrt{\ell(\ell+1)}}e^{-im\varphi_\mathbf{k}}\\
&\quad\times\Bigl[\sqrt{(\ell+m)(\ell-m+1)}Y_{\ell,m-1}(\theta^\phi_\mathbf{k},0)-\sqrt{(\ell-m)(\ell+m+1)}Y_{\ell,m+1}(\theta^\phi_\mathbf{k},0)\Bigr]\\
&=-\frac{4\pi e^{-im\varphi_\mathbf{k}}}{\sqrt{\ell(\ell+1)}}\frac{\partial Y_{\ell m}(\theta^\phi_\mathbf{k},0)}{\partial\theta}.\end{split}\end{equation}
This, together with Eq.~\eqref{PropProj}, proves the second part of Eq.~\eqref{Proj1}.

We now move to the calculation of $\bra{\ell,m,\text{M},\text{reg}}\mathbf{k},\TM,\phi\rangle$. We have
\begin{equation}\begin{split}&\exp(i\mathbf{K}^\phi\cdot\mathbf{R})\mathbf{F}^{(\text{reg})*}_{\ell m\text{M}}(\mathbf{R})\cdot\hat{\bbm[\epsilon]}_\TM^\phi(\mathbf{k},\omega)=\exp[iKR\sin\theta^\phi_\mathbf{k}\sin\theta\cos(\varphi-\varphi_\mathbf{k})]\exp[iKR\cos\theta^\phi_\mathbf{k}\cos\theta]\\
&\quad\times\frac{i^{-\ell}}{\sqrt{\ell(\ell+1)}}\Bigl[-\frac{im}{\sin\theta}\Bigl(\sin\theta^\phi_\mathbf{k}\sin\theta+\cos\theta^\phi_\mathbf{k}\cos\theta\cos(\varphi-\varphi_\mathbf{k})\Bigr)Y_{\ell m}(\theta,0)+\cos\theta^\phi_\mathbf{k}\sin(\varphi-\varphi_\mathbf{k})\frac{\partial Y_{\ell m}(\theta,0)}{\partial\theta}\Bigr]j_\ell(KR)e^{-im\varphi}.\end{split}\end{equation}
We follow the same steps as before, using twice Eqs.~\eqref{RelY} and \eqref{dY} and once Eq.~\eqref{Intej}, and we conclude
\begin{equation}\bra{\ell,m,\text{M},\text{reg}}\mathbf{k},\TM,\phi\rangle=-\frac{4\pi ime^{-im\varphi_\mathbf{k}}}{\sqrt{\ell(\ell+1)}\sin\theta^\phi_\mathbf{k}}Y_{\ell m}(\theta^\phi_\mathbf{k},0),\end{equation}
which, together with Eq.~\eqref{PropProj}, proves the first part of Eq.~\eqref{Proj1}. Associating $p=1$ ($p=2$) to TE (TM), and $P=1$ ($P=2$) to E (M) we can write
\begin{equation}\label{Gen1}\bra{\ell,m,P,\text{reg}}\mathbf{k},p,\phi\rangle=-\frac{4\pi i^{p-1}e^{-im\varphi_\mathbf{k}}}{\sqrt{\ell(\ell+1)}}\Bigl(\frac{m}{\sin\theta^\phi_\mathbf{k}}\Bigr)^{\delta_{pP}}\Bigl(\frac{\partial }{\partial\theta}\Bigr)^{1-\delta_{pP}}Y_{\ell m}(\theta^\phi_\mathbf{k},0).\end{equation}

\subsection{Projection of a spherical wave on plane waves}

The complementary problem we need to solve is the decomposition of a spherical wave in plane waves. In particular, we only need the decomposition of outgoing spherical waves, i.e. the ones involving only the Hankel functions $h_\ell^{(1)}$. We write this decomposition in the form
\begin{equation}\label{Decomp2}
\mathbf{F}^{(\text{out})}_{\ell mP}(\mathbf{R})=\sum_p\int\frac{d^2\mathbf{k}}{(2\pi)^2}e^{i\mathbf{K}^\phi\cdot\mathbf{R}}\bra{\mathbf{k},p,\phi}\ell,m,P,\text{out}\rangle\hat{\bbm[\epsilon]}_p^\phi(\mathbf{k},\omega).
\end{equation}
We want to make use of the following decomposition \cite{BobbertPhysica86}
\begin{equation}\label{Weyl}h^{(1)}_\ell(KR)Y_{\ell m}(\theta,\varphi)=\frac{2\pi i^{-\ell}}{K}\int\frac{d^2\mathbf{k}}{(2\pi)^2}\frac{1}{k_z}e^{i\mathbf{K}^\phi\cdot\mathbf{R}}Y_{\ell m}(\theta^\phi_\mathbf{k},\varphi_\mathbf{k}).\end{equation}

Taking the curl of both sides of Eq.~\eqref{Decomp2} we have, after simple manipulations,
\begin{equation}
\mathbf{F}^{(\text{out})}_{\ell mP}(\mathbf{R})=-\sum_p\int\frac{d^2\mathbf{k}}{(2\pi)^2}e^{i\mathbf{K}^\phi\cdot\mathbf{R}}\bra{\mathbf{k},S(p),\phi}\ell,m,S(P),\text{out}\rangle i(-1)^p\hat{\bbm[\epsilon]}_p^\phi(\mathbf{k},\omega),
\end{equation}
from which we deduce
\begin{equation}\label{PropProj2}
\bra{\mathbf{k},p,\phi}\ell,m,P,\text{out}\rangle=-i(-1)^p\bra{\mathbf{k},S(p),\phi}\ell,m,S(P),\text{out}\rangle.
\end{equation}

The first step is the projection of the electric field associated with the M polarization (see Eq.~\eqref{MN2}) on the two unit vectors associated to TE and TM respectively. Using Eq.~\eqref{UnVect} we obtain easily
\begin{equation}\label{FlmM}\begin{split}\mathbf{F}^{(\text{out})}_{\ell m\text{M}}(\mathbf{R})&=F^{(\text{out})\phi,\text{TE}}_{\ell m\text{M}}(\mathbf{R})\hat{\bbm[\epsilon]}_\text{TE}^\phi(\mathbf{k},\omega)+F^{(\text{out})\phi,\text{TM}}_{\ell m\text{M}}(\mathbf{R})\hat{\bbm[\epsilon]}_\text{TM}^\phi(\mathbf{k},\omega),\\
F^{(\text{out})\phi,\text{TE}}_{\ell m\text{M}}(\mathbf{R})&=\frac{i^\ell}{\sqrt{\ell(\ell+1)}}h^{(1)}_\ell(KR)\Bigl[\cos\theta\sin(\varphi-\varphi_\mathbf{k})\frac{im}{\sin\theta}Y_{\ell m}(\theta,\varphi)-\cos(\varphi-\varphi_{\mathbf{k}})\frac{\partial Y_{\ell m}(\theta,\varphi)}{\partial\theta}\Bigr],\\
F^{(\text{out})\phi,\text{TM}}_{\ell m\text{M}}(\mathbf{R})&=\frac{i^\ell}{\sqrt{\ell(\ell+1)}}h^{(1)}_\ell(KR)\Bigl[\Bigl(\sin\theta^\phi_\mathbf{k}\sin\theta+\cos\theta^\phi_\mathbf{k}\cos\theta\cos(\varphi-\varphi_\mathbf{k})\Bigr)\frac{im}{\sin\theta}Y_{\ell m}(\theta,\varphi)\\
&\,+\cos\theta^\phi_\mathbf{k}\sin(\varphi-\varphi_\mathbf{k})\frac{\partial Y_{\ell m}(\theta,\varphi)}{\partial\theta}\Bigr].\end{split}\end{equation}
We have, using Eqs.~\eqref{RelY} and \eqref{dY},
\begin{equation}\begin{split}&F^{(\text{out})\phi,\text{TE}}_{\ell m\text{M}}(\mathbf{R})=\frac{i^\ell}{\sqrt{\ell(\ell+1)}}\frac{h^{(1)}_\ell(KR)}{2}\\
&\times\Bigl[-i\sin(\varphi-\varphi_\mathbf{k})\Bigl(e^{-i\varphi}\sqrt{(\ell-m)(\ell+m+1)}Y_{\ell,m+1}(\theta,\varphi)+e^{i\varphi}\sqrt{(\ell+m)(\ell-m+1)}Y_{\ell,m-1}(\theta,\varphi)\Bigr)\\
&-\cos(\varphi-\varphi_{\mathbf{k}})\Bigl(e^{-i\varphi}\sqrt{(\ell-m)(\ell+m+1)}Y_{\ell,m+1}(\theta,\varphi)-e^{i\varphi}\sqrt{(\ell+m)(\ell-m+1)}Y_{\ell,m-1}(\theta,\varphi)\Bigr)\Bigr]\\
&=\frac{i^\ell}{\sqrt{\ell(\ell+1)}}\frac{h^{(1)}_\ell(KR)}{2}\\
&\,\times\Bigl(-e^{-i\varphi_\mathbf{k}}\sqrt{(\ell-m)(\ell+m+1)}Y_{\ell,m+1}(\theta,\varphi)+e^{i\varphi_\mathbf{k}}\sqrt{(\ell+m)(\ell-m+1)}Y_{\ell,m-1}(\theta,\varphi)\Bigr),\end{split}\end{equation}
and
\begin{equation}\begin{split}&F^{(\text{out})\phi,\text{TM}}_{\ell m\text{M}}(\mathbf{R})=\frac{i^\ell}{\sqrt{\ell(\ell+1)}}ih^{(1)}_\ell(KR)\Biggl[m\sin\theta^\phi_\mathbf{k}Y_{\ell m}(\theta,\varphi)\\
&-\frac{1}{2}\cos\theta^\phi_\mathbf{k}\Bigl[\cos(\varphi-\varphi_\mathbf{k})\Bigl(e^{-i\varphi}\sqrt{(\ell-m)(\ell+m+1)}Y_{\ell,m+1}(\theta,\varphi)+e^{i\varphi}\sqrt{(\ell+m)(\ell-m+1)}Y_{\ell,m-1}(\theta,\varphi)\Bigr)\\
&\,+i\sin(\varphi-\varphi_\mathbf{k})\Bigl(e^{-i\varphi}\sqrt{(\ell-m)(\ell+m+1)}Y_{\ell,m+1}(\theta,\varphi)-e^{i\varphi}\sqrt{(\ell+m)(\ell-m+1)}Y_{\ell,m-1}(\theta,\varphi)\Bigr)\Bigr]\Biggr]\\
&=\frac{i^\ell}{\sqrt{\ell(\ell+1)}}ih^{(1)}_\ell(KR)\Bigl[m\sin\theta^\phi_\mathbf{k}Y_{\ell m}(\theta,\varphi)\\
&-\frac{1}{2}\cos\theta^\phi_\mathbf{k}\Bigl(e^{-i\varphi_\mathbf{k}}\sqrt{(\ell-m)(\ell+m+1)}Y_{\ell,m+1}(\theta,\varphi)+e^{i\varphi_\mathbf{k}}\sqrt{(\ell+m)(\ell-m+1)}Y_{\ell,m-1}(\theta,\varphi)\Bigr)\Bigr].\end{split}\end{equation}
Finally, using \eqref{Weyl} together with Eqs.~\eqref{RelY} and \eqref{dY} we conclude
\begin{equation}\begin{split}\bra{\mathbf{k},\text{TE},\phi}\ell,m,\text{M},\text{out}\rangle&=-\frac{2\pi e^{im\varphi_\mathbf{k}}}{Kk_z\sqrt{\ell(\ell+1)}}\frac{\partial Y_{\ell m}(\theta^\phi_\mathbf{k},0)}{\partial\theta},\\
\bra{\mathbf{k},\text{TM},\phi}\ell,m,\text{M},\text{out}\rangle&=\frac{2\pi ime^{im\varphi_\mathbf{k}}}{Kk_z\sqrt{\ell(\ell+1)}\sin\theta^\phi_\mathbf{k}}Y_{\ell m}(\theta^\phi_\mathbf{k},0).\end{split}\end{equation}

This, together with Eq.~\eqref{PropProj}, proves Eq.~\eqref{Proj2}. Associating $p=1$ ($p=2$) to TE (TM), and $P=1$ ($P=2$) to E (M) we can write
\begin{equation}\label{Gen2}\bra{\mathbf{k},p,\phi}\ell,m,P,\text{out}\rangle=-\frac{2\pi i^{1-p}e^{im\varphi_\mathbf{k}}}{Kk_z\sqrt{\ell(\ell+1)}}\Bigl(\frac{m}{\sin\theta^\phi_\mathbf{k}}\Bigr)^{\delta_{pP}}\Bigl(\frac{\partial }{\partial\theta}\Bigr)^{1-\delta_{pP}}Y_{\ell m}(\theta^\phi_\mathbf{k},0).\end{equation}

\subsection{Sphere reflection matrix in plane waves and comparison with previous results}\label{RSPW}

Here we compute the matrix elements of the sphere reflection operator in the plane-wave basis
\begin{equation}\label{RS}\bra{\mathbf{k},p}\mathcal{R}_S^\phi\ket{\mathbf{k}',p'},\end{equation}
which is useful to double-check our results for the various matrix elements we have derived in this paper.
In fact, these matrix elements should satisfy the general reciprocity relations and must provide the correct small-sphere (Rayleigh) limit, which can be connected
to the atomic reflection matrix operator.

By inserting in Eq.~\eqref{RS} two closure relations in spherical waves and by exploiting Eq.~\eqref{DefRS}
\begin{equation}\bra{\mathbf{k},p}\mathcal{R}_S^\phi\ket{\mathbf{k}',p'}=\sum_{\ell=1}^{+\infty}\sum_{m=-\ell}^\ell\sum_{P\in\{\text{E},\text{M}\}}\bra{\mathbf{k},p,\phi}\ell,m,P,\text{out}\rangle\bra{\ell,m,P,\text{out}}\mathcal{R}_S^\phi\ket{\ell,m,P,\text{reg}}\langle \ell,m,P,\text{reg}\ket{\mathbf{k}',p',-\phi}.\end{equation}
Using Eqs.~\eqref{Gen1}, \eqref{DefRS} and \eqref{Gen2} we have

\begin{equation}\label{RSPWexp}\begin{split}\bra{\mathbf{k},p}\mathcal{R}_S^\phi\ket{\mathbf{k}',p'}&=\sum_{\ell=1}^{+\infty}\sum_{m=-\ell}^\ell\sum_{P\in\{\text{E},\text{M}\}}\frac{8\pi^2i^{p'-p}r_{lP}}{\ell(\ell+1)k_z}\frac{m^{\delta_{pP}+\delta_{p'P}}K^{\delta_{pP}+\delta_{p'P}-1}}{k^{\delta_{pP}}\bigl(k'\bigr)^{\delta_{p'P}}}e^{im(\varphi_\mathbf{k}-\varphi_{\mathbf{k}'})}\\
&\,\times\Bigl[\Bigl(\frac{\partial }{\partial\theta}\Bigr)^{1-\delta_{pP}}Y_{\ell m}(\theta^\phi_\mathbf{k},0)\Bigr]\Bigl[\Bigl(\frac{\partial }{\partial\theta}\Bigr)^{1-\delta_{p'P}}Y_{\ell m}(\theta^{-\phi}_{\mathbf{k}'},0)\Bigr].\end{split}\end{equation}

It is worth stressing that the sphere reflection matrix in plane waves derived from the matrix elements presented in \cite{CanaguierDurandPRA10} is related to the one given in Eq.~\eqref{RSPWexp} by the relation
\begin{equation}\Bigl(\bra{\mathbf{k},p}\mathcal{R}_S^\phi\ket{\mathbf{k}',p'}\Bigr)_\text{previous}=\sqrt{\frac{k_z}{k'_z}}\bra{\mathbf{k},p}\mathcal{R}_S^\phi\ket{\mathbf{k}',p'}.\end{equation}
Thus, the matrix elements used in \cite{CanaguierDurandPRA10} coincide with the correct ones only for $\mathbf{k}=\mathbf{k}'$. This discrepancy between the two results does not affect the results for the Casimir energy in the sphere-plane configuration discussed in \cite{CanaguierDurandPRA10}, for which only matrix elements with $\mathbf{k}=\mathbf{k}'$ are relevant. On the other hand, when more general matrix elements with any $\mathbf{k}$ and $\mathbf{k}'$ are needed, Eq.~\eqref{RSPWexp} should be used. In the rest of this Appendix, we show that Eq.~\eqref{RSPWexp} satisfies reciprocity relations (see Sec.~\ref{recrel}), as well as reproduces the reflection matrix of an atom in the limit of small radius (see Sec.~\ref{atom}).

\subsection{Reciprocity relations}\label{recrel}

We now want to prove that the sphere reflection matrix satisfies the reciprocity relation \cite{MessinaPRA11}
\begin{equation}\label{Rec}k_z\bra{\mathbf{k},p}\mathcal{R}_S^\phi\ket{\mathbf{k}',p'}=(-1)^{p+p'}k'_z\bra{-\mathbf{k}',p'}\mathcal{R}_S^\phi\ket{-\mathbf{k},p},\end{equation}
which for our purpose becomes
\begin{equation}\begin{split}&k_z\sum_{\ell=1}^{+\infty}\sum_{m=-\ell}^\ell\sum_{P\in\{\text{E},\text{M}\}}\bra{\mathbf{k},p,\phi}\ell,m,P,\text{out}\rangle\bra{\ell,m,P,\text{out}}\mathcal{R}_S^\phi\ket{\ell,m,P,\text{reg}}\langle \ell,m,P,\text{reg}\ket{\mathbf{k}',p',-\phi}\\
&=(-1)^{p+p'}k'_z\sum_{\ell=1}^{+\infty}\sum_{m=-\ell}^\ell\sum_{P\in\{\text{E},\text{M}\}}\bra{-\mathbf{k}',p',\phi}\ell,-m,P,\text{out}\rangle\bra{\ell,-m,P,\text{out}}\mathcal{R}_S^\phi\ket{\ell,-m,P,\text{reg}}\langle \ell,-m,P,\text{reg}\ket{-\mathbf{k},p,-\phi},\end{split}\end{equation}
where in the second sum we changed $m$ in $-m$. It is then sufficient to prove that
\begin{equation}\begin{split}r_1&=k_z\bra{\mathbf{k},p,\phi}\ell,m,P,\text{out}\rangle\langle \ell,m,P,\text{reg}\ket{\mathbf{k}',p',-\phi}\\
&=(-1)^{p+p'}k'_z\bra{-\mathbf{k}',p',\phi}\ell,-m,P,\text{out}\rangle\langle \ell,-m,P,\text{reg}\ket{-\mathbf{k},p,-\phi}=r_2.\end{split}\end{equation}
To this aim we first observe from Eq.~\eqref{Angles} that
\begin{equation}\label{PropAngles}\begin{split}\sin\theta^\phi_\mathbf{k}&=\sin\theta^{-\phi}_\mathbf{k}=\sin\theta^\phi_{-\mathbf{k}},\\
\cos\theta^\phi_\mathbf{k}&=-\cos\theta^{-\phi}_\mathbf{k}=\cos\theta^\phi_{-\mathbf{k}},\\
\sin\varphi_\mathbf{k}&=-\sin\varphi_{-\mathbf{k}},\\
\cos\varphi_\mathbf{k}&=-\cos\varphi_{-\mathbf{k}},\end{split}\end{equation}
Using Eqs.~\eqref{Gen1} and \eqref{Gen2} we have
\begin{equation}r_1=k_z\Bigl(\frac{2\pi i^{1-p}}{Kk_z}\frac{4\pi i^{p'-1}}{\ell(\ell+1)}\Bigr)\Bigl(\frac{m}{\sin\theta^\phi_\mathbf{k}}\Bigr)^{\delta_{pP}}\Bigl(\frac{m}{\sin\theta^{-\phi}_{\mathbf{k}'}}\Bigr)^{\delta_{p'P}}e^{im(\varphi_\mathbf{k}-\varphi_{\mathbf{k}'})}\Bigl[\Bigl(\frac{\partial}{\partial\theta}\Bigr)^{1-\delta_{pP}}Y_{\ell m}(\theta^\phi_\mathbf{k},0)\Bigr]\Bigl[\Bigl(\frac{\partial}{\partial\theta}\Bigr)^{1-\delta_{p'P}}Y_{\ell m}(\theta^{-\phi}_{\mathbf{k}'},0)\Bigr],\end{equation}
and
\begin{equation}\begin{split}r_2&=k'_z(-1)^{p+p'}\Bigl(\frac{2\pi i^{1-p'}}{Kk'_z}\frac{4\pi i^{p-1}}{\ell(\ell+1)}\Bigr)\Bigl(\frac{-m}{\sin\theta^\phi_{-\mathbf{k}'}}\Bigr)^{\delta_{p'P}}\Bigl(\frac{-m}{\sin\theta^{-\phi}_{-\mathbf{k}}}\Bigr)^{\delta_{pP}}e^{-im(\varphi_{-\mathbf{k}'}-\varphi_{-\mathbf{k}})}\\
&\,\times\Bigl[\Bigl(\frac{\partial}{\partial\theta}\Bigr)^{1-\delta_{p'P}}Y_{\ell,-m}(\theta^\phi_{-\mathbf{k}'},0)\Bigr]\Bigl[\Bigl(\frac{\partial}{\partial\theta}\Bigr)^{1-\delta_{pP}}Y_{\ell,-m}(\theta^{-\phi}_{-\mathbf{k}},0)\Bigr].\end{split}\end{equation}
Using Eq.~\eqref{PropAngles} and the properties of spherical harmonics
\begin{equation}\begin{split}r_2&=(-1)^{p+p'}\Bigl(\frac{8\pi^2i^{p-p'}}{K\ell(\ell+1)}\Bigr)\Bigl(\frac{m}{\sin\theta^{-\phi}_{\mathbf{k}'}}\Bigr)^{\delta_{p'P}}\Bigl(\frac{m}{\sin\theta^\phi_\mathbf{k}}\Bigr)^{\delta_{pP}}e^{im(\varphi_\mathbf{k}-\varphi_{\mathbf{k}'})}\\
&\,\times\Bigl[\Bigl(\frac{\partial}{\partial\theta}\Bigr)^{1-\delta_{p'P}}Y_{\ell m}(\theta^{-\phi}_{\mathbf{k}'},0)\Bigr]\Bigl[\Bigl(\frac{\partial}{\partial\theta}\Bigr)^{1-\delta_{pP}}Y_{\ell m}(\theta^\phi_\mathbf{k},0)\Bigr],\end{split}\end{equation}
which coincides indeed with $r_1$ since $i^{p'-p}=(-1)^{p+p'}i^{p-p'}$.
\end{widetext}

\clearpage

\subsection{Limit of small radius: atomic reflection matrix}\label{atom}

We want to show here that in the limit of small radius of the sphere we are able to recover analytically the reflection matrix of an atom obtained within the dipolar approximation in \cite{MessinaPRA09}. It is well known that in the case of a small sphere the term $\ell=1$ gives the main contribution. Furthermore, the matrix elements with $P=\text{E}$ are proportional to the Mie coefficient $a_1$, which dominates over $b_1$, present in the case $P=\text{M}$. The Mie coefficient $a_1$ can be approximated as
\begin{equation}\begin{split}\bra{1,m,\text{E}}\mathcal{R}_S^\phi\ket{1,m,\text{E}}&=a_1\simeq\frac{2i}{3}(KR)^3\frac{\varepsilon(\omega)-1}{\varepsilon(\omega)+2}\\
&=\frac{2iK^3}{3}\frac{\alpha(\omega)}{4\pi\varepsilon_0},\end{split}\end{equation}
where we have introduced the polarizability of the sphere. We then have
\begin{equation}\begin{split}\bra{\mathbf{k},p}\mathcal{R}_S^\phi\ket{\mathbf{k}',p'}&=\frac{2iK^3}{3}\frac{\alpha(\omega)}{4\pi\varepsilon_0}\sum_{m=-1}^1\bra{\mathbf{k},p,\phi}1,m,\text{E},\text{out}\rangle\\
&\qquad\times\langle1,m,\text{E},\text{reg}\ket{\mathbf{k}',p',-\phi}.\end{split}\end{equation}
We have
\begin{equation}\begin{split}
\bra{\mathbf{k},\text{TM},\phi}1,0,\text{E},\text{out}\rangle&=\sqrt{\frac{3\pi}{2}}\frac{i^{-1}\sin\theta^\phi_\mathbf{k}}{Kk_z},\\
\bra{\mathbf{k},\text{TM},\phi}1,\pm1,\text{E},\text{out}\rangle&=\pm\frac{\sqrt{3\pi}i^{-1}\cos\theta^\phi_\mathbf{k}}{2Kk_z}e^{\pm i\varphi_\mathbf{k}},\end{split}\end{equation}
\begin{equation}\begin{split}
\bra{\mathbf{k},\text{TE},\phi}1,0,\text{E},\text{out}\rangle&=0,\\
\bra{\mathbf{k},\text{TE},\phi}1,\pm1,\text{E},\text{out}\rangle&=\frac{\sqrt{3\pi}}{2Kk_z}e^{\pm i\varphi_\mathbf{k}},\end{split}\end{equation}
and
\begin{equation}\begin{split}
\bra{1,0,\text{E},\text{reg}}\mathbf{k},\TM,\phi\rangle&=\sqrt{6\pi}i\sin\theta^\phi_\mathbf{k},\\
\bra{1,\pm1,\text{E},\text{reg}}\mathbf{k},\TM,\phi\rangle&=\pm\sqrt{3\pi}i\cos\theta^\phi_\mathbf{k}e^{\mp i\varphi_\mathbf{k}}.\end{split}\end{equation}
\begin{equation}\begin{split}
\bra{1,0,\text{E},\text{reg}}\mathbf{k},\TE,\phi\rangle&=0,\\
\bra{1,\pm1,\text{E},\text{reg}}\mathbf{k},\TE,\phi\rangle&=\sqrt{3\pi}e^{\mp i\varphi_\mathbf{k}},\end{split}\end{equation}
Using these relations, it can be shown that
\begin{equation}\begin{split}\sum_{m=-1}^1&\bra{\mathbf{k},p,\phi}1,m,\text{E},\text{out}\rangle\langle1,m,\text{E},\text{reg}\ket{\mathbf{k}',p',-\phi}\\
&=\frac{3\pi}{Kk_z}\bigl(\hat{\bbm[\epsilon]}_p^\phi(\mathbf{k},\omega)\cdot\hat{\bbm[\epsilon]}_{p'}^{-\phi}(\mathbf{k}',\omega)\bigr),\end{split}\end{equation}
thus
\begin{equation}\bra{\mathbf{k},p}\mathcal{R}_S^\phi\ket{\mathbf{k}',p'}=\frac{2\pi iK^2}{k_z}\frac{\alpha(\omega)}{4\pi\varepsilon_0}\bigl(\hat{\bbm[\epsilon]}_p^\phi(\mathbf{k},\omega)\cdot\hat{\bbm[\epsilon]}_{p'}^{-\phi}(\mathbf{k}',\omega)\bigr),\end{equation}
which is the reflection matrix of an electrically polarizable atom located in the origin \cite{MessinaPRA09}.

\end{document}